\documentclass[6pt,dvipsnames,twocolumn]{article}
\usepackage{amsmath}
\usepackage{graphicx}
\usepackage{geometry}
\usepackage{sectsty}
\usepackage{xcolor}
\usepackage{etoolbox}
\usepackage{amsmath}
\usepackage{subcaption}
\usepackage[labelformat=parens,labelsep=quad,skip=3pt]{caption}
\usepackage{titlesec}
\usepackage{url}
\RequirePackage{hyperref}

\patchcmd{\thebibliography}{\section*{\refname}}{}{}{}
\graphicspath{ {./images/} }
\titlespacing\section{0pt}{12pt plus 4pt minus 2pt}{0pt plus 2pt minus 2pt}
\titlespacing\subsection{0pt}{12pt plus 4pt minus 2pt}{0pt plus 2pt minus 2pt}
\titlespacing\subsubsection{0pt}{12pt plus 4pt minus 2pt}{0pt plus 2pt minus 2pt}

\begin{document}

\twocolumn[{\centering{\huge Network memory in the movement of hospital patients carrying drug-resistant bacteria \par}\vspace{4ex}
	{Ashleigh C. Myall$^{1,2}$, Robert L. Peach$^{1}$, Andrea Y. Wei{\ss}e$^{2,3}$, Frances Davies$^{2,4}$, Siddharth Mookerjee$^{4}$, \\ Alison Holmes$^{2,4}$ and Mauricio Barahona$^{1}$
	\par
	}\vspace{2ex}
	    {\small
$^{1}$ Department of Mathematics, Imperial College London, London, UK.}\\
{\small
$^{2}$ Department of Infectious Disease, Imperial College London, London, UK.}\\
{\small
$^{3}$ Current address: School of Informatics, University of Edinburgh, Scotland, UK.}\\
{\small
$^{4}$ Imperial College Healthcare NHS Trust, London, UK.}
	
	\vspace{2ex}\today\par\vspace{4ex}}
	
{\centering\bfseries Abstract\par\vspace{1ex}}
Hospitals constitute highly interconnected systems that bring into contact an abundance of infectious pathogens and susceptible individuals, thus making infection outbreaks both common and challenging. In recent years, there has been a sharp incidence of antimicrobial-resistance amongst healthcare-associated infections, a situation now considered endemic in many countries. Here we present network-based analyses of a data set capturing the movement of patients harbouring drug-resistant bacteria across three large London hospitals. We show that there are substantial memory effects in the  movement of hospital patients colonised with drug-resistant bacteria. Such memory effects break first-order Markovian transitive assumptions and substantially alter the conclusions from the analysis, specifically on node rankings and the evolution of diffusive processes.  We capture variable length memory effects by constructing a lumped-state memory network, which we then use to identify overlapping communities of wards. We find that these communities of wards display a quasi-hierarchical structure at different levels of granularity which is consistent with different aspects of patient flows related to hospital locations and medical specialties.

\par\vspace{5ex}]

\section*{Introduction}

Antimicrobial resistance (AMR) poses one of the greatest risks to human health \cite{prestinaci2015}. Currently, around 700,000 people worldwide die from infections with resistant pathogens each year, and this number is estimated to rise to up to 10 Million by 2050  \cite{interagency_coordination_group_on_antimicrobial_resistance_no_2019, ONeil2016}. Hospitals and other healthcare facilities act as key vectors for the spread of AMR through healthcare-associated infections (HAI) \cite{struelens1998}. Persistent colonisation of hospital patients and the networked nature of hospital processes underlying patient mobility will likely cause AMR to remain prevalent \cite{pastor2001_end}. Several factors moreover exacerbate the spread of AMR in healthcare facilities, including the selective pressures generated by increased antimicrobial usage, and the large pool of vulnerable patients, who are more susceptible to infections \cite{who2002}. The need for infection prevention and control (IPC) can therefore not be understated.

Understanding the transmission dynamics of AMR promises valuable insights to improve IPC strategies. Key to these measures will be the analysis of patient pathways capturing the movement of patients carrying AMR during their hospital stay. Like many real-world systems, healthcare facilities have complex structure, which when ignored can limit the insights into the underlying dynamic processes. In this study we focus on mapping the movement pathways of patients known to carry antimicrobial-resistant bacteria onto physical structures of the hospitals. Specifically, we focus on patients colonised with Carbapenemase-producing Enterobacteriaceae (CPE). CPE is a particularly concerning form of AMR that confers resistance to carbapenems, broad-spectrum antibacterials often used as last-line antibiotics. CPE infections have recently seen a global surge amongst HAIs~\cite{bonomo2018,logan2017}.

Networks provide an powerful formalism to analyse the movement of patients in hospitals. Nodes typically represent physical locations within the hospital, such as wards, and edges represent the flow of patients between these locations, with edge weights encoding the volume of patient flow from one location to another. To facilitate analysis, we can aggregate the movements of individual patients into probabilities of transitioning between hospital wards~\cite{donker2012,bean2017}. Typically, patient trajectories are broken down into individual transitions between wards: first, the number of transitions between each ward is summed across all patients and subsequently, for each ward the sum of all out-going transitions is normalised to one. The constructed network may then be interpreted as a first-order Markov model, where a random walker transitions with a probability proportional to the observed outflow volume from the current node to others in the network~\cite{salnikov2016}. 

This dynamical assumption, whilst useful because of its simplicity and ease of implementation, is however limited by the assumption that transitions between nodes are independent of prior nodes within the patient pathway. Previous studies have indeed shown that first-order Markovian dynamics are not sufficient to fully model network dynamics of disease propagation \cite{Robert2001,Pastor-Satorra2001}. Akin to shipping trajectories or passenger movement between airports, patient movement in hospitals tends to follow particular patterns dictated by medical or operational constraints. In particular, it is plausible that patient trajectories could bear ‘memory’, that is, a subsequent move depends on several or all previous locations visited, and not solely on the current location leading to transitive dependence in the data.

Introduced by Shannon~\cite{Shannon1948}, higher-order memory models have shown relevance across a number of applications, and a wide range of real world movement data \cite{Song2010,kareiva1983,chierichetti2012,singer2014,gonzalez2008,heath2008} including several epidemiological data sets~\cite{balcan2011,poletto2013}. Ignoring such transitive dependencies and modelling patient movement via memory-less, first-order Markov models can distort both network topology and conclusions on the underlying process~\cite{mucha2010}. Despite the clear importance of transitive dependence, to date we only find one study \cite{palla2018} of hospital patient movement accounting for these relationships, and none when looking at AMR across healthcare facilities. Hence, in this study we investigate evidence for and implications of \emph{transitive dependencies} in the movement patterns of hospital patients colonised with a CPE by including \emph{memory} in our network models.

\begin{figure}[!htb]
    \centering
    \includegraphics[width=0.45\textwidth]{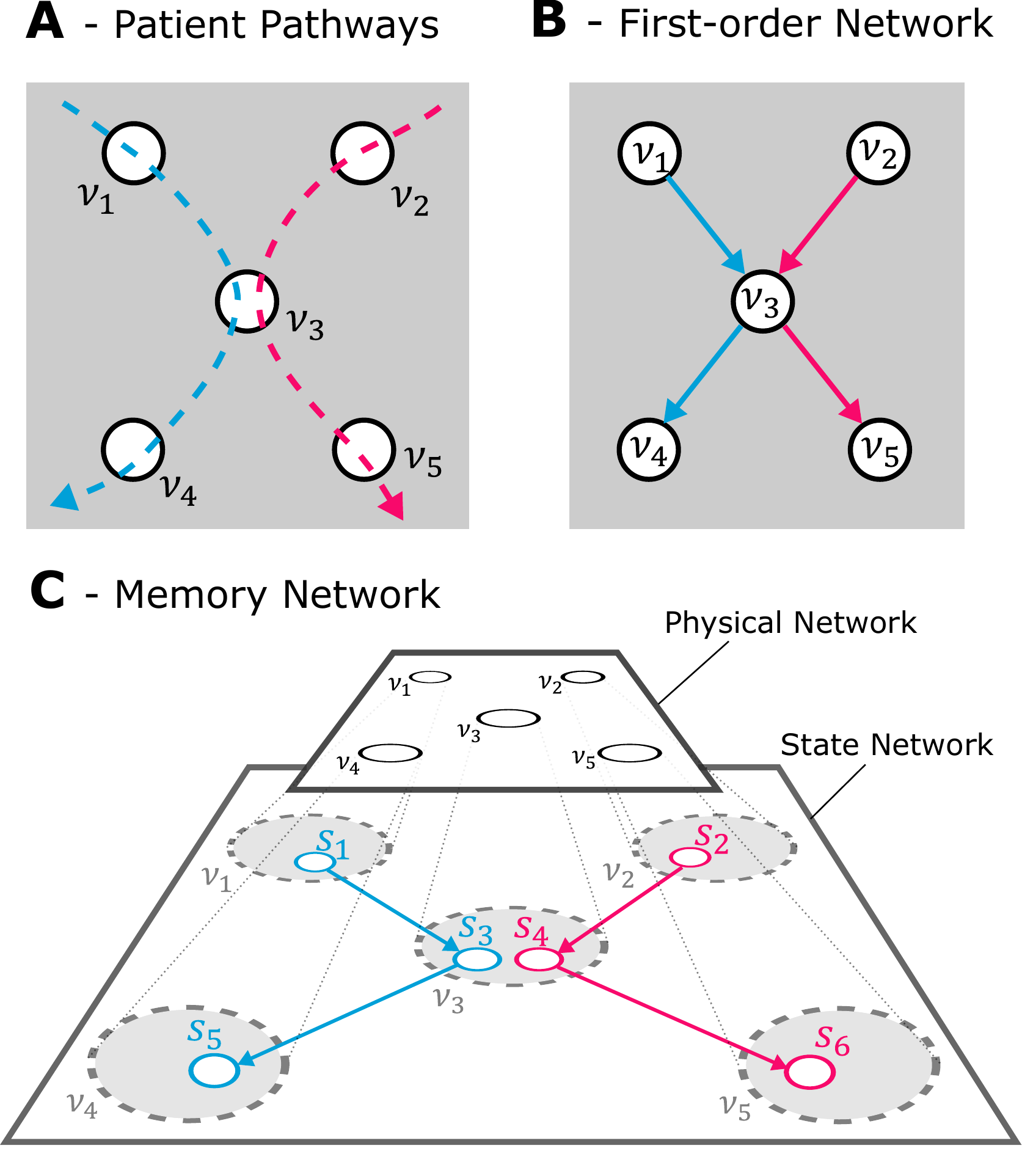}
    \caption{Illustration of transitive dependence encoded into memory network. (A) Two sets of typical patient pathways, largely independent, but passing through the same ward as an intermediate point in their pathways. (B) First-order representation of A without any memory (C) Memory network representation A, whereby a physical node network maps to state nodes, which encode transitive dependence of the patient pathways and constrain a random walkers movement.}
    \label{fig:illistrative_figure}
\end{figure}

To model these effects, we use \emph{memory networks}, which encode the memory of individual trajectories into higher-order transitive relationships, and which have successfully been used to investigate transitive dependence in pathway data \cite{lambiotte2019}. To provide some intuition behind memory networks, consider a simple example of a small network of a hospital with five wards where the patients can follow one of two possible routes between the wards, and the two routes share one common node (Figure \ref{fig:illistrative_figure}A). A first-order (memory-less) network model assuming full transitivity (Figure \ref{fig:illistrative_figure}B) would wrongly suggest that a patient starting at $v_1$ could transition to $v_5$ with some probability, when in fact, only patients starting at $v_2$ can reach $v_5$. In a memory network (Figure \ref{fig:illistrative_figure}C) these transitive dependencies are captured by abstracting away from a network of physical nodes to a higher-order networks of state nodes representing the possible dynamical states of the system  (i.e., the sequence of hospital wards visited up to a given memory)~\cite{edler2017}. Specifically, in a memory network each state represents a pathway of length $k-1$, whereby higher-order state networks increase the length $k$ of the pathways captured by each state. This state network can be thought of as an additional layer of information still bound to the physical network since each state node is assigned to a physical node. The state network thus acts to constrain how a random walker transitions between physical nodes. These higher-order network abstractions lend themselves to learning tasks that can pinpoint key properties underlying the dynamical process. In the case of HAIs, this can offer insight into more accurate patterns in the movement of infected patients otherwise lost in a network model that assumes full transitivity.

Below we present the analysis of patients pathways confirmed to be colonised with CPE. We begin by presenting our data and a description of the hospital network. We then present evidence for memory within patient pathways by contrasting models constructed with and without memory. Finally, we construct a lumped state memory network, which captures transitive dependence and removes redundancy. We carry out multiscale community detection on this network, and present the resultant communities, highlighting specific wards and specialities that are important across different regions of the network.

\section*{Results}

\subsection*{Data}

\begin{figure*}[!htb]
    \centering
    \includegraphics[width=0.8\textwidth]{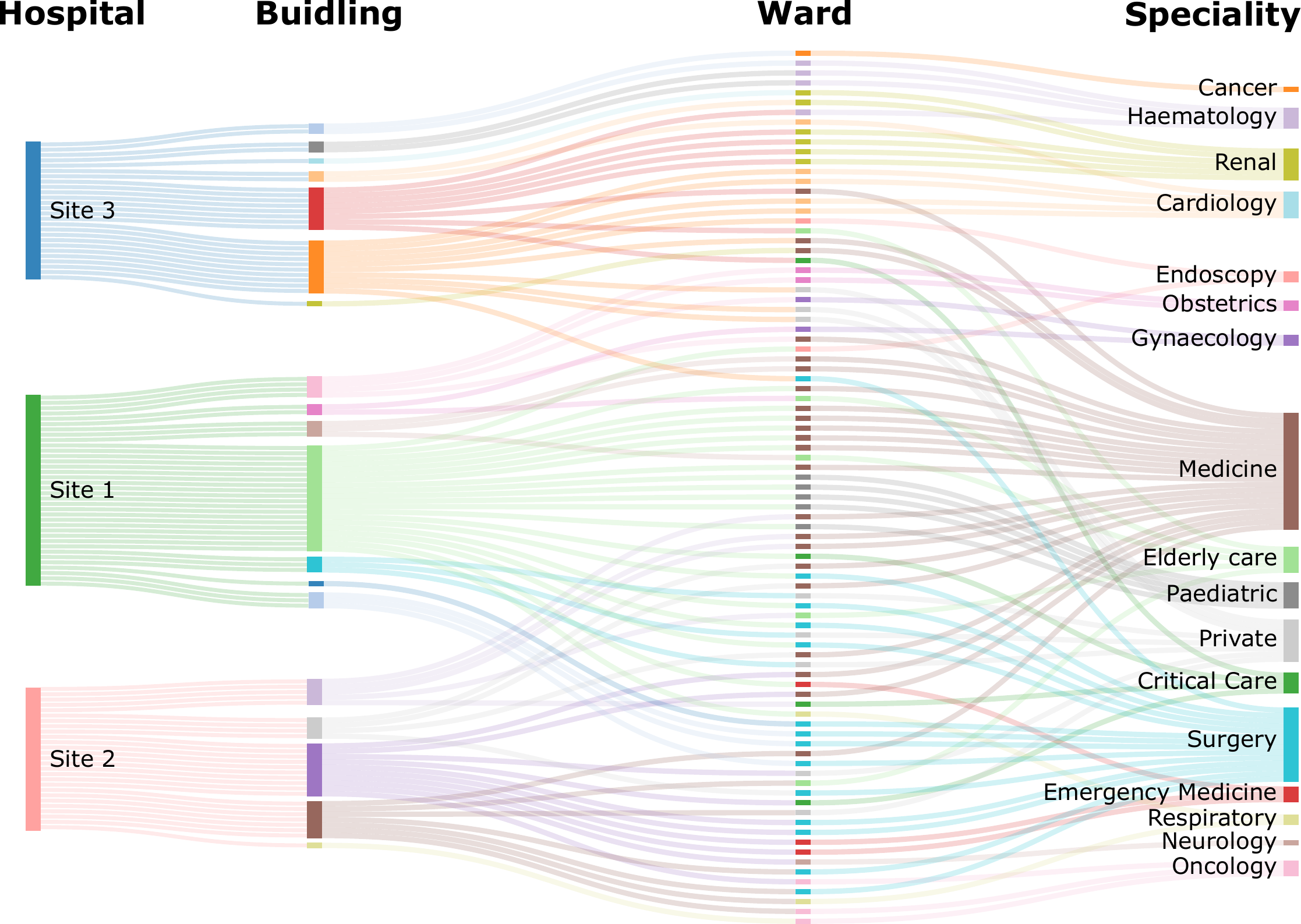}
    \vspace*{5mm}
    \caption{Sankey diagram of Trust Structure traversed by CPE patients. Broken down by hospital site, and buildings to wards, then also broken down by speciality into wards. 
    }
    \label{fig:hospital_formal_struc}
\end{figure*}

Our analysis is based on anonymised electronic health records of patients from a large 1000-bed Trust of London teaching hospitals. Specifically, we used ward-level movement patterns of 967 patients who tested positive for CPE over a period of two years between 2018 and 2020. We focused on the subset of 526 patients who moved between at least two wards during their hospital stay for a total of 1958 transitions between 96 hospital wards.

Formally, the hospital Trust is structured around 17 \emph{specialities} and 19 \emph{buildings}, the latter belonging to three \emph{hospital sites} (Figure \ref{fig:hospital_formal_struc}). Hospital site 3 acts as a Tertiary site with only speciality wards. Whilst sites and buildings are constrained by geographical factors, specialities are defined by medical procedures and thus may overlap across sites and buildings. In fact, a number of specialities span all five hospital sites (Critical Care, Elderly Care, Medicine, Private, and Surgery). Geographical structures constrain patient movement to some extent: patients with certain co-morbidities and therapeutic requirements are likely to be constrained to a single or several specialities supporting those needs, whereas other patients can move within buildings, or between wards placed closely for logistics and ease of transfer.

\subsection*{From Patient Pathways To Network Models}

We consider the trajectories of $p$ patients. Each patient pathway as a trajectory $T_\alpha$ and the set of $\alpha=1,\ldots,p$ trajectories is $\mathcal{T} = \begin{Bmatrix} T_1,T_2,T_3,...,T_p \end{Bmatrix}$. Each $T_\alpha$ consists of a time-ordered set representing discrete movements between nodes,
\begin{equation} 
\label{eqn:transitionProbK1}
    T_\alpha = \begin{Bmatrix}{ v_i  \rightarrow v_j \rightarrow ... \rightarrow v_k}\end{Bmatrix},
\end{equation}
where each node refers to one of $N$ hospital wards $\mathcal{N} = \begin{Bmatrix} \nu_1,\nu_2,\nu_3,...,\nu_N \end{Bmatrix}$. Since these nodes represent physical locations, we will refer to them as \emph{physical nodes} to avoid confusion with \emph{state nodes}, which we introduce next. 

In order to understand the aggregate dynamics of all patients, whilst preserving memory effects in $\mathcal{T}$, we represent the trajectories as a memory network as proposed by Rosvall \emph{et al.} \cite{rosvall2014}. This way, we maintain information about physical nodes $\mathcal{N}$ whilst instilling transitive dependence in the connectivity patterns of an underlying state-network, $\mathcal{M}_k = (\mathcal{E}_k,\mathcal{S}_k)$. Here $\mathcal{E}_k$ is the set of edges that link the set of state nodes $\mathcal{S}_k$ that capture higher-order memory of order $k$ \cite{edler2017}.

A memory network of order $k = 1$,  $\mathcal{M}_1$, represents a system with zero memory, where the movement of a random walker only depends on its current location. In this special case, the state network $\mathcal{M}_1$ is equivalent to an aggregated physical network $G=(\mathcal{E},\mathcal{N})$, and the set of states directly maps to the set of physical nodes, i.e., $\mathcal{S}_1 = \mathcal{N}$. The edge weights $w_{ij}$ conforming the set $\mathcal{E}$ in $\mathcal{M}_1$ represent the frequency of transitions between physical nodes $\nu_i$ and $\nu_j$ across the set of trajectories $\mathcal{T}$. Given $w_{ij}$, we can write the transition probability matrix $P_1$ for $\mathcal{M}_1$ as
\begin{equation} 
\label{eqn:transitionProbK2}
    p_{ij} = P(i \rightarrow j) =  \frac{w_{ij}}{ \sum\nolimits_{j} w_{ij} } .
\end{equation} 

In memory networks of higher-order, where $k>1$, state nodes represent pathways of length $k - 1$, and are no longer equivalent to the physical nodes $\mathcal{S}_k \ne \mathcal{N}$. This representation allows us to introduce the memory dependence in $\mathcal{T}$, capturing multi-step patterns of flow via the state nodes of the network~\cite{salnikov2016}.

In particular, for the second-order memory network $\mathcal{M}_2$, a state node represents a directed pathway of length one $s_j = \overrightarrow{ij}$. For two states nodes $s_j=\overrightarrow{ij}$ and $s_\ell=\overrightarrow{j
\ell}$ to be connected, a path of length two, ($\nu_i \rightarrow \nu_j \rightarrow \nu_\ell$), must occur in the set of trajectories $\mathcal{T}$. Similarly for higher-order models, edges between state nodes are weighted $w_{s_j s_l}$ and capture the number of occurrences that a transition between state nodes $s_j$ and $s_ l$ was observed in $\mathcal{T}$. Transition probabilities $P_k$ of $\mathcal{M}_k$ for any order can be derived from Equation \ref{eqn:transitionProb_ij_k} by altering the state node set $S$ to represent pathways of length $k - 1$, so
\begin{equation} 
\label{eqn:transitionProb_ij_k}
    p_{s_i s_j} = P(s_i \rightarrow s_j) =  \frac{s_{ij}}{ \sum\nolimits_{j} s_{ij} } .
\end{equation} 
Each state node can be mapped to a physical node (Figure \ref{fig:illistrative_figure}A), using an $|\mathcal{S}_k| \times N$ indicator matrix $D$, the elements of which, $D_{s\nu}  \in \begin{Bmatrix}0,1\end{Bmatrix} $, indicate the final physical node of a pathway $s$.

We first constructed a first-order memory network $\mathcal{M}_1$ that contains 96 state nodes with a one-to-one mapping to the 96 wards (physical nodes). $\mathcal{M}_1$ consists of four weakly connected components, one of which contains the majority of state nodes (87 out of 96) (Additional file 1 ). We next constructed a second-order memory network $\mathcal{M}_2$ that contains 384 state nodes, in 18 weakly connected components. Similarly, $\mathcal{M}_2$ consists of a single weakly component that contains the majority of state nodes (329 of 384). Structurally, $\mathcal{M}_1$ has a higher connectivity with a clustering coefficient of 0.287 and a diameter of 6, whereas $\mathcal{M}_2$ is more sparse with a clustering coefficient of 0.003 and a larger diameter of 31, resembling a series of connected line graphs (Additional file 1).

\subsection*{Patient trajectories break first-order dynamics}

Using random walks to reveal and probe the structure of networks has long been a foundational tool in network science~\cite{Masuda2017}. A random walk is a stochastic process which consists of a succession of random steps with no memory of its past locations; however, in a system where transitive dependence plays a important role, a purely random walk becomes inaccurate and potentially misleading. Memory networks of higher-order with $k>1$ can capture deviations from first-order transitive assumptions by constraining where a random walker can next move depending on its previous location(s). For pathway models without transitive dependence, a random walker should be no more constrained when moving from a first-order memory network, $\mathcal{M}_1$, to a second-order memory network $\mathcal{M}_2$ ~\cite{rosvall2014}. However, pathways exhibiting transitive dependence will constrain a random walker comparatively more in the second-order memory network. Here, we use the entropy rate of the random-walk to measure the uncertainty of moving between two state nodes~\cite{Shannon1948}: 
\begin{equation}
\label{eqn:entropy}
    H(S_{t+1}|S_t) =  \sum_{i,j}  \pi (i) p(i \rightarrow j)  \log p (i \rightarrow j),
\end{equation} 
where $\pi$ denotes the stationary distribution across  $\mathcal{M}$, and $p(i \rightarrow j)$ are the transition probabilities.

\begin{figure*}[!htb]
    \centering
    \includegraphics[width=0.75\textwidth]{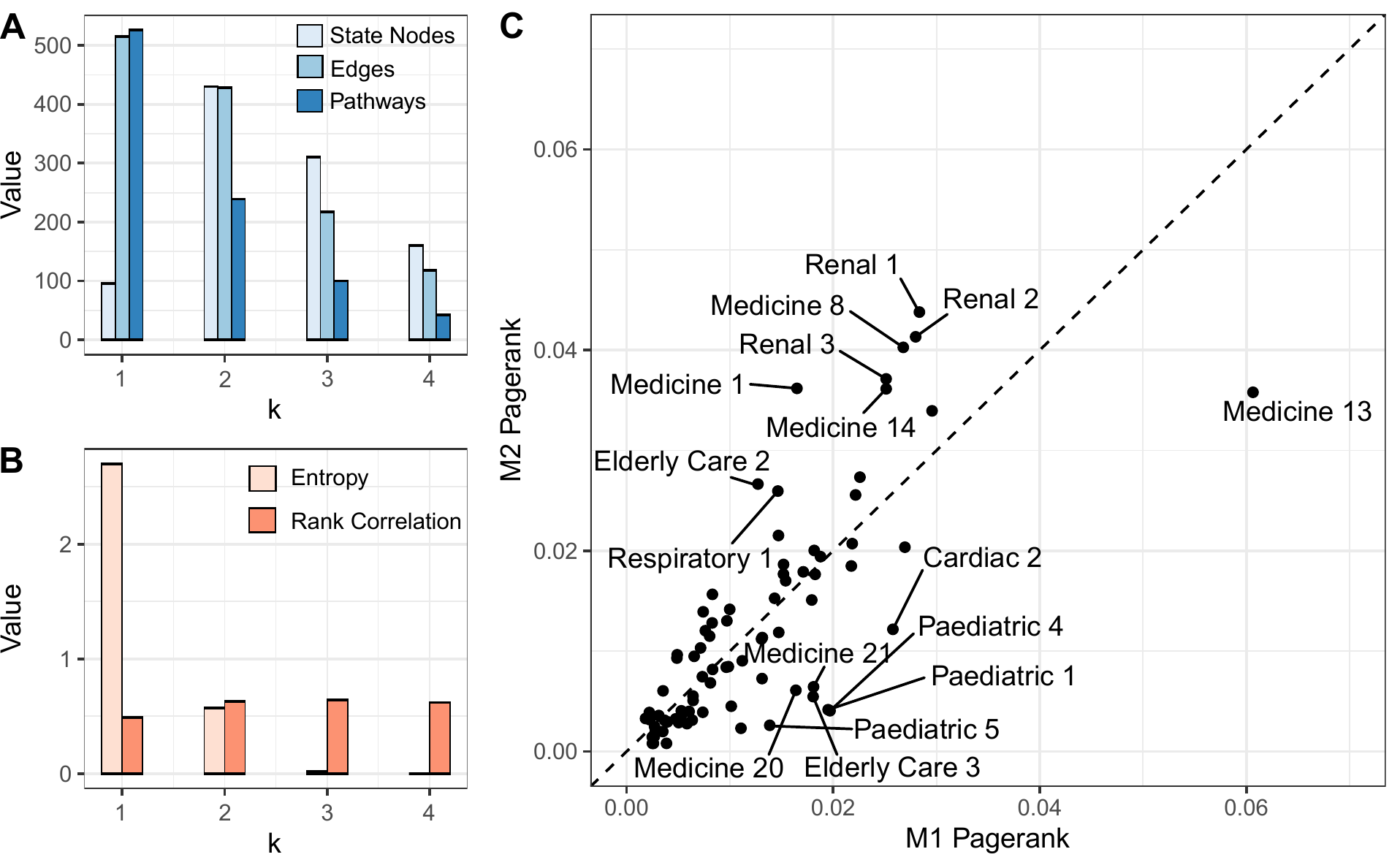}
    \vspace*{5mm}
    \caption{A comparison of the effect of memory for models of order $k$. (A) Details of network size (no. state nodes, edges) and data (no. trajectories) for $k = {1,2,3,4}$. (B) The entropy of each higher order memory network and the rank correlation between train and validation sets. (C) PageRank of wards for $\mathcal{M}_1$ and $\mathcal{M}_2$, dashed line indicating equality of PageRank.}
    \label{fig:higher_order_validation}
\end{figure*}

We constructed the memory networks $\mathcal{M}_k$ for $k = {1,2,3,4}$ (description of the number of state nodes, edges and pathways for each $\mathcal{M}_k$ is detailed in  Figure \ref{fig:higher_order_validation}A). Computing the entropy for each $\mathcal{M}_k$ we find increasing restriction of the random walk (reduced entropy) for larger $k$ (Figure \ref{fig:higher_order_validation}B). In particular, we observe a large decrease in entropy from 2.70 to 0.57 when we move from  $k=1$ to $k=2$. Patient pathways with little to no memory effect would not exhibit any large change in entropy when moving from $\mathcal{M}_1$ to $\mathcal{M}_2$ and thus our results suggest that there exist patient pathways which break first-order Markovian transitive assumptions and highlight the importance of capturing memory.

Now we must determine the optimal order $k$ for a given analysis. For small data sets, it is difficult to statistically validate whether memory networks with higher-order are relevant, given that the parameter space and complexity increases exponentially \cite{scholtes2017}. A common workaround is to use cross-validation, a model validation technique borrowed from machine learning \cite{singer2014}. In cross-validation, data is partitioned and performance is determined as an average across partitions to reduce over-fitting and selection bias \cite{arlot2010,cawley2010}. To perform cross-validation in the framework of a memory network we compute the rank orders of wards using a training set of patient pathways and then compare with the rank order of wards generated from visitation probabilities of a withheld partition of patient pathways. Similar to \emph{Rosvall et al.} \cite{rosvall2014}, we used a generalised PageRank for higher-order models where the visitation probabilities of state nodes were summed for each physical node (see methodology). The rank orders between train and test sets were compared with Kendall-Tau rank correlation \cite{kendall1938} and the results were averaged over a 5-fold cross-validation split. We found that $\mathcal{M}_2$ was more predictive of the node ranking of physical nodes than $\mathcal{M}_1$ (0.60 to 0.49) (Figure \ref{fig:higher_order_validation}B). This increased performance in $\mathcal{M}_2$ again suggests that a patient's current and previous location both affect future movement, and that accounting for this memory effect yields more accurate approximations of patient movement.

Whilst further higher-order memory effects may exist, we were unable to detect any increased predictive power beyond $k=2$ (Figure \ref{fig:higher_order_validation}B). We note that this may be due to limitations of our data; as we increase the order $k$, we must discard additional patient pathways with fewer than $k$ transitions between wards. This is evident in Figure \ref{fig:higher_order_validation}A that shows a decreasing number of pathways as we increase the $k$; and whilst the number of state nodes and edges initially increases from $k=1$ to $k=2$, as you may expect by increasing model complexity, due to the decreasing number of pathways we instead observe a decrease in the number of state nodes beyond $k>2$. Herein, to retain enough patient pathways for reliable insights, we thus shall focus on the $k=2$ memory network.

We then compared the PageRank of physical nodes (wards) between $\mathcal{M}_1$ and $\mathcal{M}_2$ (Figure \ref{fig:higher_order_validation}C). Whilst we found the PageRank of wards in $\mathcal{M}_1$ and $\mathcal{M}_2$ to be correlated (0.81 (pval$<$0.01)), there were a number of key deviations. In particular, we find three renal wards (Renal 1, 2 \& 3) with a relatively higher ranking in $\mathcal{M}_2$, indicating that CPE patients frequently visit these wards. Given that patients undergoing renal therapies are particularly at an increased risk for CPE acquisition within this hospital group \cite{otter2020}, it is no surprise we find these wards with a high PageRank in both $\mathcal{M}_1$ and $\mathcal{M}_2$. However, the \emph{higher} ranking of these Renal wards in $\mathcal{M}_2$ highlights the importance of using a constrained state node network to understand the clinical movement of these patients. Conversely, Medicine 13 was the highest ranked ward in $\mathcal{M}_1$, but was found to have a relatively lower rank in $\mathcal{M}_2$. Medicine 13 is an acute medical admissions unit, and as such acts as the entry/re-entry point for many patients to the hospital, rather than a transition ward or a ward which offers care, and whilst it plays a starting role in many patient pathways, it is seldom observed at any other point in a patients trajectory through the hospital. 

\subsubsection*{Investigating memory effects with a discrete diffusive process}

\begin{figure*}[!htb]
    \centering
    \includegraphics[width=0.99\textwidth]{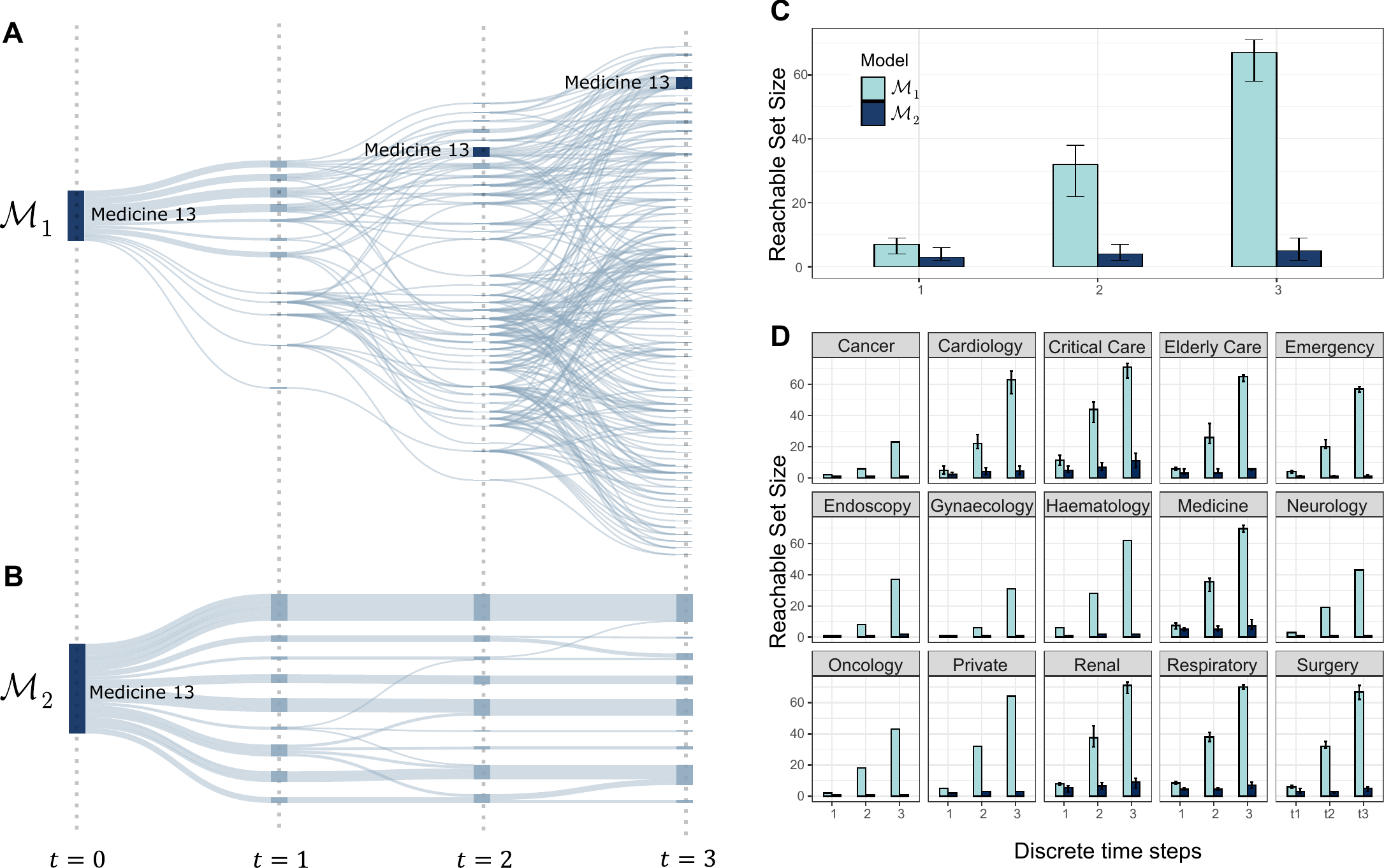}
    \vspace*{5mm}
    \caption{Memory effects on network pathways. (A\&B) A discrete random walk starting at Medicine 13 in the state networks of $\mathcal{M}_1$ and $\mathcal{M}_2$, modelled as a discrete time diffusion over the transition matrices $P_1$ and $P_2$ respectively. After a single discrete time step $t=1$ the reachable number of wards is similar between $\mathcal{M}_1$ and $\mathcal{M}_2$, however, they quickly diverge as the diffusion evolves over time and the transitive effects increase in $\mathcal{M}_1$. (C\&D)  Median size of the reachable set of wards
    for $\mathcal{M}_1$ and $\mathcal{M}_2$: (C) Overall reachability after $t$-discrete steps, and (D) reachability after $t$-discrete steps broken down by speciality (for in the largest connected component).
    }
    \label{fig:memory_affect_visual}
\end{figure*}

One way we can study the effect of memory is through the direct observation of its influence on a diffusion process starting at various points in the network~\cite{lambiotte2019}. Figure \ref{fig:memory_affect_visual} A\&B displays the evolution of a discrete-time diffusive process for $\mathcal{M}_1$ and $\mathcal{M}_2$, each encoded by their respective transition matrix $P_k$,  when injecting an impulse at a single ward (Medicine 13). At time $t=0$, the diffusive process is entirely contained within the state node(s) corresponding to Medicine 13 (for $k>1$, where physical nodes can have several state nodes, we share initial probability over states based on the frequency of pathways in $\mathcal{T}$ that $s_j = \overrightarrow{ij}$ represented). For times $t>0$ we compute the probabilities of being on a given ward at time $t$ through powers of the transition matrix $P_k^t$. 

After a single discrete step $t=1$ we find there is little effect of memory with the total number of wards reachable being similar for $\mathcal{M}_1$ and $\mathcal{M}_2$ (12 wards vs 9 wards, respectively). 
However, as we extend the diffusive process to $t=2$ and $t=3$ we find that the number of reachable wards from Medicine 13 increases rapidly for $\mathcal{M}_1$ (36 wards at $t=2$, then 71 wards at $t=3$) whereas we do not see any change in $\mathcal{M}_2$
(9 wards at $t=2$, and 9 wards at $t=3$). In fact, for $\mathcal{M}_1$ a random walk initialised at Medicine 13 can reach 71 out of the 79 wards within the largest weakly connected component in $\mathcal{T}$ after just 3 steps. This level of transitivity is not present in $\mathcal{T}$, and its absence is directly observable by looking at the restriction of flow evident in $\mathcal{M}_2$ (Figure \ref{fig:memory_affect_visual} A\&B). This difference comes from patients not starting at Medicine 13, but passing through its neighbours influencing the 2-step network transitivity. 
Interestingly, $\mathcal{M}_2$ constrained walkers such that no backtracking to Medicine 13 is possible over the first 3 discrete transitions, in contrast to $\mathcal{M}_1$, where backtracking to Medicine 13 is possible for $t>1$. In fact, using $\mathcal{M}_1$ there is a relatively large probability to revisit Medicine 13 after 2 or 3 steps ($p^2_{med13}=0.18$ and $p^3_{med13}=0.24$). Given that Medicine 13 is commonly an entry point/readmission point where patients go when waiting for diagnosis, we would expect a minimal backtracking effect in patient movement across short time frames since they move into subsequent specialities for treatment once a diagnosis is known. Hence, including memory through $\mathcal{M}_2$ better captures true patient flow. 

\subsubsection*{Forward reachability is varyingly constrained by memory}

We expanded the above framework to examine reachability across the entire network by performing the analysis for every possible starting node. For each ward, we compute the set of reachable wards after $t$ time-steps and in Figure \ref{fig:memory_affect_visual}C we display the median size of reachable sets for all wards under $\mathcal{M}_1$ and $\mathcal{M}_2$. Similar to the analysis of Medicine 13 in Figure \ref{fig:memory_affect_visual} A\&B, we find that the median size of reachable sets is relatively similar between $\mathcal{M}_1$ and $\mathcal{M}_2$ at $t=1$. However, as $t$ increases we again observe divergence in the reachable set sizes due to the significantly larger set of reachable wards in the first-order model $\mathcal{M}_1$. Indeed, after 3 time-steps only 5 wards are reachable on average under $\mathcal{M}_2$ as compared to the 79 reachable wards under $\mathcal{M}_1$. Hence $\mathcal{M}_1$ is inflating transitivity between wards and distorting the set of reachable wards for a patient through inherent ignorance of prior ward visits. We also observe that the variance of the reachable set of wards for $\mathcal{M}_1$ increases for $t=2,3$, suggesting that the importance of memory is different depending on the ward at which the diffusive process was initialised. 

To study this, we next break down wards by speciality and examine the importance of memory on the median set size of reachable wards.
Figure \ref{fig:memory_affect_visual}D summarises the size of reachable sets averaged across wards within the same speciality. We notice that specialities which are known to be well visited by CPE patients in this hospital setting (e.g., Critical Care, Renal) exhibit a comparatively larger reachability set size when compared to the aggregated view in Figure \ref{fig:memory_affect_visual}C. In contrast, specialities such as Neurology or Cancer which are less common to CPE patients exhibit a relatively lower reachability. These different reachabilities between specialities could be the consequence of two mechanisms: (1) the different roles specialities play within the network and their transitivity by CPE patient trajectories, and (2), that memory effects may vary in different areas of the network, i.e. the extent to which a previous ward determines a patients next move. Hence, it may be optimal to construct a ‘hybrid’ of $\mathcal{M}_1$ and $\mathcal{M}_2$ which incorporates many of the desirable memory effects in $\mathcal{M}_2$, but simplifies parts of the model where greater transitivity is in fact present.

\subsection*{Reducing complexity using state node lumping}

Given a large set of trajectories, the problem arises that state node networks $\mathcal{M}_k$ can become increasingly large and often duplicate or contain redundant information. In the case of patient trajectories, not all hospital pathways may exhibit memory effects in equal measure. Variable-length Markov models, pioneered by Rissanen \cite{rissanen1983} alleviate some of these issues by introducing a `lumping' step in which `redundant' states are merged, thus enabling models to capture variable lengths of memory and remove model redundancy \cite{jaaskinen2014,buhlmann1999}. Remembering that in memory networks, state nodes are assigned to physical nodes, we will often find several state nodes that are connecting the same physical nodes just via different edges. There is no need for this repetition and therefore here we focused on lumping state nodes within the same physical node to form so called `meta-state nodes' or `lumped nodes' which also benefit from preserving the physical network structure~\cite{lambiotte2019}. For each lumped node, we reassemble all connections between two states nodes such that weighting and connectivity are preserved \cite{edler2017}. In effect, `lumping' nodes retains \emph{relevant} and distinct patterns of transitive dependence in the original pathways; however, for our purposes it also serves to 'de-sparsify' $\mathcal{M}_2$, improving its practicality and making it useful for subsequent learning tasks that assume greater connectivity.

In our approach, we lump together state nodes based on the similarity of visitation probabilities computed from a discrete diffusive process encoded in the state transition matrix $P_k$ over $t$-steps. Existing node lumping methodologies use a 1-step random walk to identify state nodes that have similar connectivity within the network\cite{edler2017,persson2016}. Here, however, we extend this approach to $t$-steps to identify similarity across a greater network locality. Using an agglomerative clustering method on the discrete diffusive process, we can then identify state nodes with similar connectivity, and if both are members of the same physical node they can be lumped together\cite{hastie2009} (for a detailed explanation see methods).

To what extent should we lump state nodes together? At one extreme, we have the state node network $\mathcal{M}_k$ without any lumping and at the other extreme we have the physical node network where every state node has been lumped together within its respective physical node. We want to identify an optimal lumping, comfortably between the two extremes, that retains transitive dependence but removes redundant or duplicated information. The resulting lumped network is denoted $\hat{\mathcal{M}}_k$. In order to quantitatively determine the optimal lumping, we used `ground-truth' community structures such as buildings, specialities, and hospital sites and compared these annotations with the results of community detection on the lumped network $\hat{\mathcal{M}}_k$. Whilst these structures do not fully constrain patient movement and therefore cannot provide an exact ground truth, there does exist a correlation with patient movement. We hypothesised that the optimal lumping would be found at the elbow of a fitness curve generated from the ability to detect known hospital structures in community structures, thus providing a trade-off between model accuracy and simplicity. Accordingly, we found that a lumping rate of $r=0.35$ gave the optimal lumped model (Additional file 3) .

\begin{figure}[!htb]
    \centering
    \includegraphics[width=0.45\textwidth]{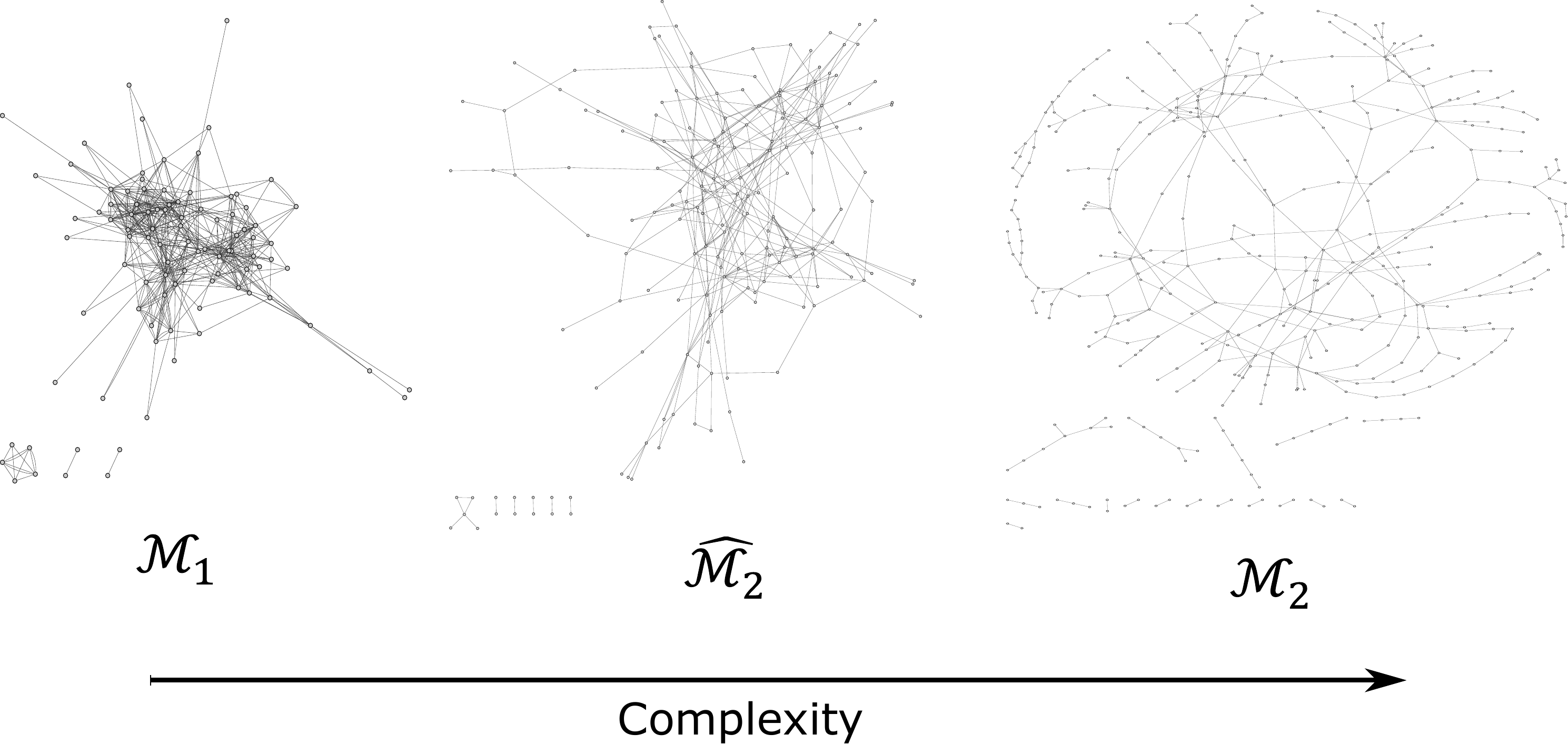}
    \vspace*{5mm}
    \caption{From first-order network to second-order network and everything between. The first order network  $\mathcal{M}_1$ (left), the lumped second-order network $\hat{\mathcal{M}_2}$ (middle), and the second-order state node network $\mathcal{M}_{2}$ (right) ordered by scale of model complexity.}
    \label{fig:m1_m2_m2Lumped_networks}
\end{figure}

The lumped network $\hat{\mathcal{M}_2}$ contains 171 state nodes across 7 weakly connected components. Similar to the state node networks $\mathcal{M}_1$ and $\mathcal{M}_2$, we found a large weakly connected component that contained the majority of state nodes (156 out of 171) (Figure \ref{fig:m1_m2_m2Lumped_networks}). Aside from visually appearing to exist in a state between $\mathcal{M}_1$ and $\mathcal{M}_2$, both its clustering coefficient (0.054) and network diameter (11) sat comfortably between $\mathcal{M}_1$ and $\mathcal{M}_2$, serving to validate its balance of complexity, connectivity, and higher-order dependencies. Note that unlike $\mathcal{M}_2$, the lumped state network $\hat{\mathcal{M}_2}$ no longer resembles a series of lines graphs, and hence provides a more practical structure over which to apply community detection.

\subsection*{Community detection reveals overlapping clusters of wards common to distinct pathways}

By constraining a walkers movement within the connectivity patterns of $\mathcal{M}_k$, for $k > 1$, we can identify communities within $\mathcal{M}_k$ that conserve flow from a dynamical perspective. Given that $\mathcal{M}_k$ is composed of state nodes, the memory-dependent structure $\mathcal{C}$ will provide network partitions that shed light into community structure. Here we use Markov Stability (MS), a quasi-hierarchical community detection algorithm that identifies regions within a network in which a diffusive process becomes transiently constrained~\cite{delvenne2008}. MS exploits diffusion dynamics over an underlying graph structure to reveal multi-scale community organisation and their stability across time scales (see methods for a more detailed introduction).

\subsubsection*{The quasi-hierarchical community structure of the wards}

\begin{figure}[!htb]
    \centering
    \includegraphics[width=0.46\textwidth]{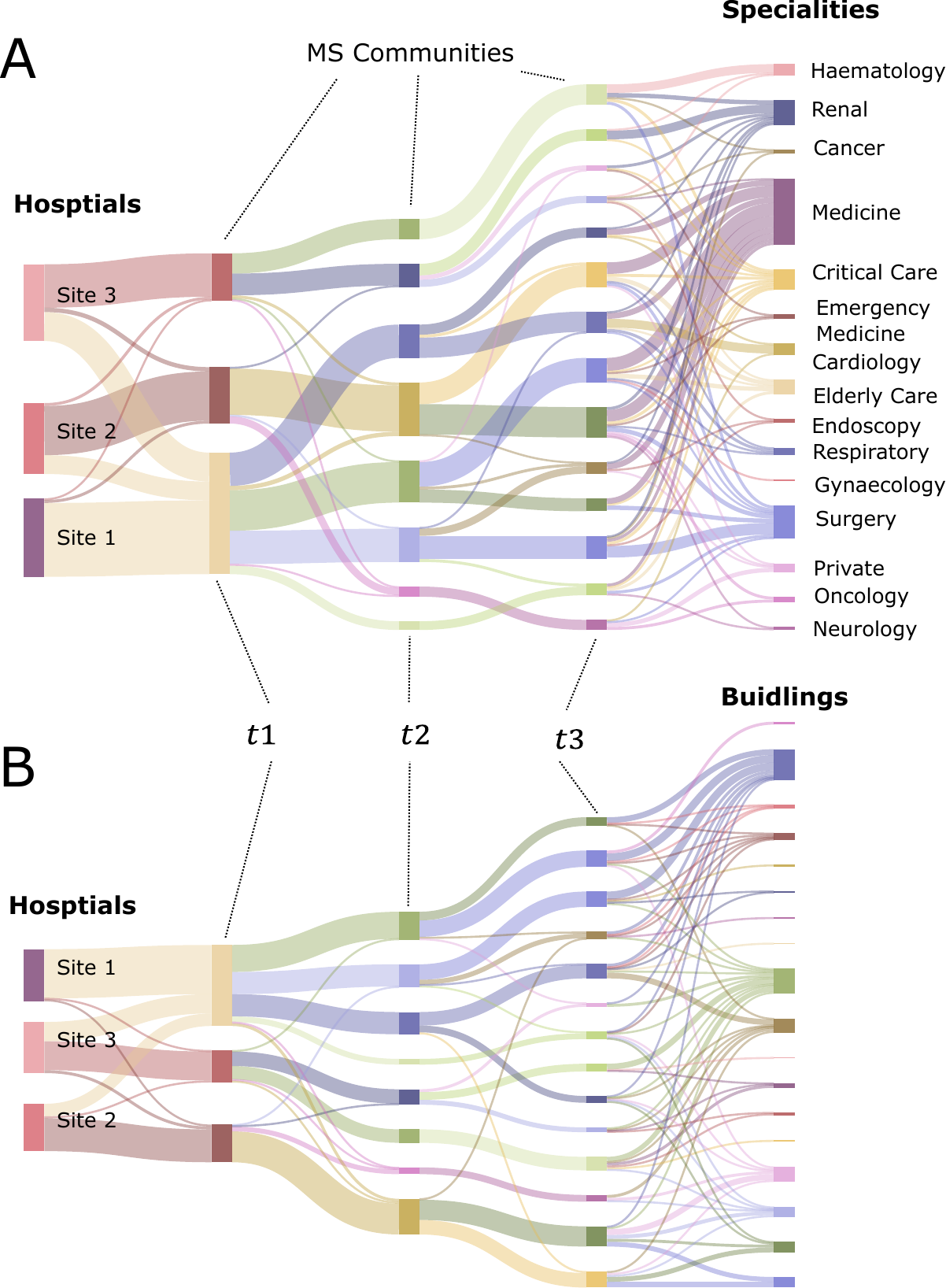}
    \vspace*{5mm}
    \caption{Hierarchical breakdown of Markov Stability communities for three chosen points in Markov time (optimal partitions in Markov time chosen for their robustness after a more thorough Markov Stability analysis, see Additional file 4) and their relations to Hospital sites for coarse partitions, and then their relations at granular partitions to (A) Specialities and (B) Buildings.}
    \label{fig:community_detect_MS}
\end{figure}

Continuing with the lumped state network $\hat{\mathcal{M}_2}$, we apply MS and in Figure \ref{fig:community_detect_MS} we show an apparent hierarchy of state node assignments to community partitions across Markov time $t$. We selected three points across Markov time ($t_1$,$t_2$,$t_3$) that exhibited robust community partitions(Additional file 5). At longer time scales MS reveals coarser community partitions which show significant correspondence to hospital sites (Figure \ref{fig:community_detect_MS}). Specifically, at $t_3$ each cluster in the 3-way partition strongly corresponds to one of the three hospital sites. If we extend to even longer $t$ we identify a 2-way partition where two hospitals are grouped almost exclusively into a single community (Additional file 7). Notably, the hospital with wards grouped separately is the Tertiary site within the hospital trust which consists of speciality wards and appears to share fewer patients with the other two hospitals.

Moving towards shorter $t$ within the MS analysis, which are expected to identify more granular structures of patient flow, we identity sub-structures largely contained \emph{within} hospital sites, which overlap to a lesser extent between hospital sites. In some cases, these confer to buildings (we find 10 buildings that are over-represented in clusters at $t_1$), in other cases these confer to specialities (we find 7 specialities over-represented in clusters at $t_1$). Focusing initially on speciality, we find three specialities (Haematology, Cardiology, and Renal) that are over-represented within separate communities suggesting they are have a high degree of within speciality patient movement (Figure \ref{fig:community_detect_MS}A). However, as we increase $t$ to reveal coarser partitions we see the more granular communities combine, bringing together previously distinct specialities such as Haematology or Renal into coarser partitions with other specialities, highlighting the zooming affect of MS as we change the $t$ at which communities are observed. However, it is clear that the community structure is not entirely defined by specialities and the physical constraints imposed by buildings, hospitals, and common movement patterns play a significant role and result in our observed communities. (Figure \ref{fig:community_detect_MS}B).
Given that the majority of patients will move between specialities at some point during their journey through the hospital, it is expected that communities would not correspond exactly to ward specialities. A number of specialities will service several different groups of patients such as those in Medicine, a general class of ward that often takes admissions, or Critical Care, which can service patients from any given ward if they deteriorate fast enough. Notably, we find that Wards both in the Medicine and Critical Care specialities can be found within 10 different communities at $t_1$, additionally, Surgery another department services multiple other wards, can be found in 9 different communities.

\subsubsection*{Overlapping community assignments}

Community detection generally focuses on finding disjoint communities, however, \emph{multiple community memberships} is a well observed  phenomena, whereby a given node may have multiple functions that it shares with different groups of nodes \cite{xie2013}. Understanding that we are essentially clustering wards based on the movement patterns of patients, it is likely that different cohorts of CPE patients (e.g. with different comorbidities) have overlapping pathways. For instance, different cohorts of patients still require a set of common services and hence visit an overlapping set of wards (e.g. for admission, surgery, critical care, or renal dialysis). This phenomenon is well captured by memory networks, standard methods of community detection applied across the state network are able to reveal overlapping communities of nodes on the physical network. Additionally, the notation of granularity introduced by MS adds an interesting dimension to this problem, whereby the degree to which wards overlap communities can depend on the point Markov time. We can thus identify hospital wards which persistently overlap multiple communities across both granular and coarse time scales. These wards are of particular interest when developing Infection Prevention and Control strategies as they can play the role of network \emph{bridges} and potential transmission hotspots.

At the most granular time scales, we find 48 wards with multiple community assignments (Additional file 8). With increasing Markov time the total number of overlapping wards decreases; however, there exist several wards which are persistently overlap communities. We find 4 Renal wards and a single Elderly Care ward which have membership within each community of the 2-way coarse partition. Despite disappearing in the very coarser 2-way partition after $t>12$, Critical Care, Medicine, and Surgery, as well as a single Elderly care ward also overlapped between communities. Since the most coarse partitions strongly corresponded to non-specialist hospital, and specialist hospital sites, it is likely that Critical Care and the Elderly care wards still play a strong connective role within connecting the two non-specialist hospital Sites 1\&2.

\subsection*{Identifying the most central wards}

In the previous section we identified nodes that were assigned to multiple communities, highlighting their critical role in the pathways of multiple cohorts of patients with differing patterns, and prior, we examined the PageRank of wards to identify their importance in $\mathcal{M}_1$ and $\mathcal{M}_{2}$.

For a more complete examination of ward importance, and investigation into $\hat{\mathcal{M}_2}$, we use Multiscale Centrality (MSC), that enables us to identify nodes that are central at different scales within the network \cite{PhysRevResearch.2.033104}. Following the same approach to compute centrality of the physical nodes, we compute MSC for each state node and then compute the sum of state node centrality across each physical node to generate a value of MSC for each ward.

\begin{figure}[!htb]
    \centering
    \includegraphics[width=0.45\textwidth]{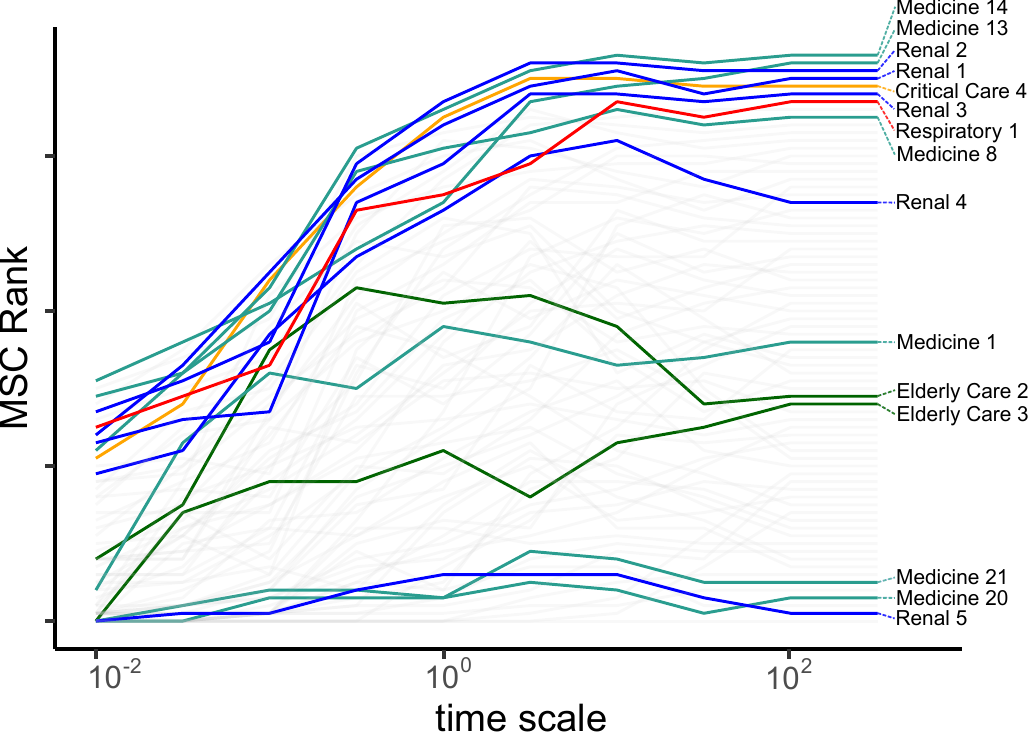}
    \vspace*{5mm}
    \caption{Multiscale centralities ranking of $\hat{\mathcal{M}_2}$ across time. The several wards annotated are those with largely different PageRanks in the comparison of $\mathcal{M}_1$, $\mathcal{M}_2$. At short Markov timescales, some nodes will have not been assigned a Multiscale centrality value and so will not yet be ranked.}
    \label{fig:msc}
\end{figure}

Figure \ref{fig:msc} shows the results of MSC computed for $\hat{\mathcal{M}_2}$. We find several wards that are central at all scales, implying that they are both highly connected locally (short time scales), and also important as global connectors/bridges (long time scales). Both Medicine 13 and 14 appear as central at all time-scales; Medicine 13 and 14 are both admission and readmission points into the hospital, where patients will be first identified as positive for CPE, and where they will return if readmitted. Additionally, we find 4 renal wards are central at all scales. Interestingly, we also find wards which vary considerably in their importance across time-scales; Elderly Care 2 seems well connected locally, but at longer time scales its becomes comparatively less important.

\section*{Conclusions}

Analysing a large set of patient pathways, we showed that the movement of hospitalised patients colonised with CPE displays substantial memory effects. This means that ward transitions depend on previously visited wards. Memory effects were evident from the difference between the node rankings of different order models, as well as the statistics of a diffusion process on the resulting network models. Notably, memory increased the probabilities of visiting wards known to be commonly visited by CPE patients (e.g. Renal) and decreased the probabilities of visiting wards less common amongst CPE patients (e.g. Paediatric). Memory also greatly affected local reachability; for example, the memory-less first-order model, wrongly implied almost any ward could be reached from any other wards within three discrete time steps. Our work thus showed that not accounting for pathway ‘memory’ can mislead both the importance of hospital wards and knowledge about how patients move throughout healthcare networks. These insights into the constraints of the movement of CPE patients can aid infection prevention and control to prevent transmission to uncolononised patients. 

Models with memory have substantially larger parameter space. We therefore simplified the memory model by constructing a hybrid 'lumped' memory network. The latter retains the effect of distinct memory present in the patient trajectories but removes redundant or duplicate information. In this context, we extended previous work on lumping in memory networks in two ways: firstly, by defining a state node feature vector, which allowed state nodes to be compared and lumped into meta nodes based on longer random-walks; secondly, we proposed that lumping could be optimised by using prior knowledge with known communities which partially constrain patient pathways.

The lumped memory network then formed the basis of our subsequent investigation using community detection to reveal communities of movement within our healthcare network. To this end, we used prior knowledge, including the hospital structure or specialities with noisy signals, to optimise the rates of lumping based on Markov Stability. As a result, we can highlight pathway clusters with higher-order memory, and identify wards that occur across multiple pathway communities. Particularly, we found that community overlaps identified wards that are visited by virtually all CPE patients (Renal wards), or wards visited commonly by the general patient population (Medicine, Surgery, and Critical Care wards). Notably, there may be some ward selection bias here, due to the nature of the medical conditions of the patients who specifically attend the renal wards, which mean they have an increased risk of CPE carriage; although this was not studied for this analysis, connections by medical diagnoses could inform future work. The communities of CPE patient movement we identified divided the hospital sites quasi-hierarchically into sub-communities of wards that share patient flow. There was some correlation between community structures and known structures, such as hospital buildings or specialities, however, communities likely result from common pathways specific to certain cohorts of CPE patients amongst this hospital group.

Our study highlights the role of memory in patient pathways. Most current analyses of patient pathways assume memoryless-ness. Here, however, we showed that ignoring memory may wrongly identify potential hubs of disease transmission. This in turn would mislead efforts to prevent infection of the general patient population. Our lumped memory networks therefore provide a framework for future patient-pathway analyses to improve containment of CPE and may as well be applied to inform infection prevention and control of other HAIs. 

\section*{Methods}

\subsection*{Higher-order PageRank}\label{sec:pagerank}

PageRank is a measure of node importance or centrality within a network based on the incoming edges~\cite{page1999}.
To obtain \emph{Higher-order} PageRank we follow the derivation presented by \emph{Rosval et al.} in ~\cite{rosvall2014}. PageRank is essentially computing the visitation probabilities to nodes over a network, determined by connectivity and weighting of those connections. In the context of a memory network, one can simply derive PageRank over the underlying state network for a model of arbitrary order $k$, and project the visitation probabilities back onto the physical nodes.

Firstly, we define the probability of finding a random walker on a given state node $s$ at time $t+1$ as

\begin{equation}
\label{eqn:prob_nodej}
    P(s_j; t+1) =  \sum_{s_i} P(s_i;t)p(s_i \rightarrow s_j),
\end{equation} where as before a state confers to a pathway of length $k$ and transition probabilities are encoded by the transition matrix $P$.

Now, for any order $k$ the higher-order generlisation of PageRank is simply the stationary solution to equation \ref{eqn:prob_nodej}:
\begin{equation}
     \pi (s_j) =  \sum_{i}  \pi (s_i) p(s_i  \rightarrow s_j).
\end{equation}

With $\pi (s_j)$ it is then trivial to return the physical node PageRank by summing over a physical nodes states:

\begin{equation}
     \pi (k) =  \sum_{j}  \pi (s_j) =  \sum_{k}  \pi (s_j).
\end{equation}

\subsection*{State lumping on local connectivity}

\begin{figure*}[h]
    \centering
\includegraphics[width=0.90\textwidth]{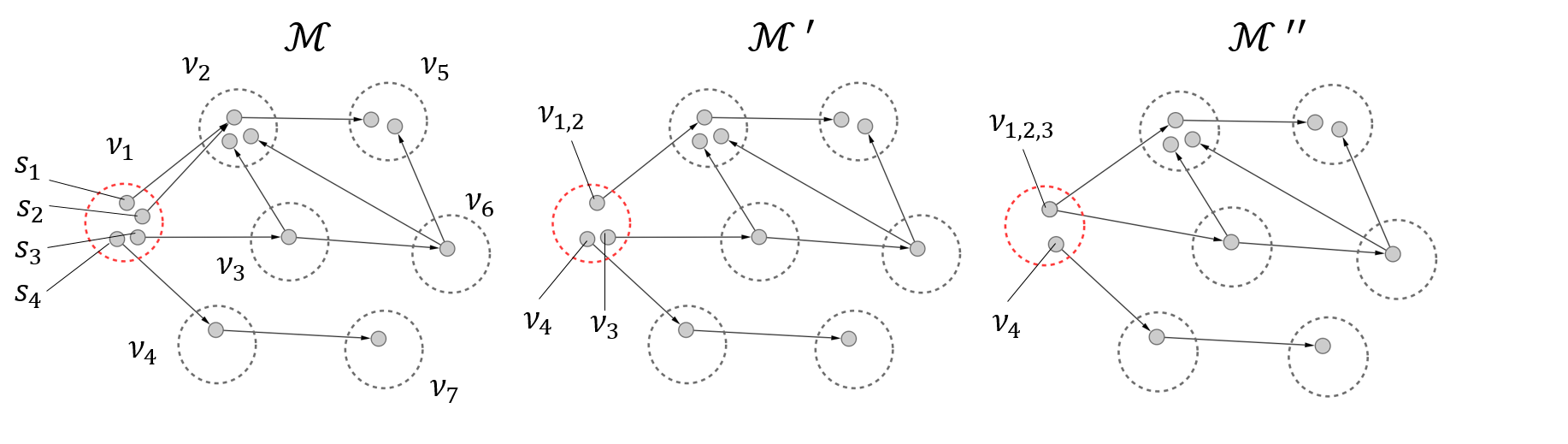}
    \caption{State Lumping Example for a single physical node ($\nu_1$). Two possible  lumpings $\mathcal{M}'$ and $\mathcal{M}''$ are visualised here over the state nodes (grey nodes) with the physical node mapped over (dotted circles surrounding state nodes). Here each lump merges the two most similar state nodes based on feature vectors capturing local visitation probabilities of k=2 network steps.}
    \label{fig:figure_1}
\end{figure*}

Given a large set of trajectories, the problem arises that state node networks $\mathcal{M}_k$ can become very large and often contain redundancies. Not all pathways exhibit full transitive dependence, so it can often be desirable to reduce the model complexity by lumping together redundant state nodes. Redundancy of state nodes can lead to over-fitting when a physical node contains a number of similar states. Hence, we focus on lumping states nodes within the same physical node, forming so called `meta state nodes' which also benefit from preserving the physical network structure~\cite{lambiotte2019}. For each lump, we reassemble all connections between two states nodes such that transition probabilities and connectivity are preserved \cite{edler2017}. In effect, `lumping' state nodes together reduces the model complexity whilst retaining the transitive dependence of the original pathways.

In our approach, we lump together state nodes based on the similarity of visitation probabilities of the physical nodes. To do this we use the $S \times S$ state transition matrix $P$ over $k$-steps and then sum the probabilities over the state nodes that compose each physical node. 
In the construction of $P$ we add weighted self loops equivalent to a nodes total outflow weight $w_{s_is_i} = \sum\nolimits_{s_i } w_{ s_i s_j }$ to derive $P^\prime$ with Equation \ref{eqn:transitionProbK1}. This self loop conserves local flow across $P^\prime$, emphasising local connectivity when we subsequently determine distances across $X$. 

We define the state node to physical node transition matrix $X$ as the visitation probabilities of each state node to each physical node over $k$-steps, $X = P^{k}D$ , where $P$ is the state node transition matrix and $D$ is the $S \times N$ state node to physical node indicator matrix. Each entry $x_{ij}$ corresponds to the probability of transitioning from state node $i$ to physical node $j$ and thus provides a mapping from the higher order state node network to the physical node network. Here, we set $k=3$ to incorporate a slightly larger range of local connectivity than previous works that use $k=1$~\cite{edler2017,persson2016}.

State nodes with similar local connectivity will exhibit similar probability distributions on the physical node network, therefore we can compute a similarity matrix between state nodes by computing the Wasserstein distance \cite{villani2008} between vector rows of $X$ which measures the distance for moving from one probability distribution to another. The similarity matrix was subsequently clustered using an agglomerative clustering method for lumping state nodes within physical node \cite{hastie2009}.

In order to control the lumping of state nodes we employed a clustering rate $r$, which sets the number of final lumped state nodes that should be constructed for each physical node after completion of the lumping procedure. For example, lets consider a scenario where we have two physical nodes, one of which is composed of 10 state nodes and the second is composed of 20 state nodes. If we set the lumping rate $r=0.2$, then after lumping the first physical node would have 2 lumped nodes after the procedure whereas the second physical node would have 4 lumped nodes. Increasing the lumping rate to $r=0.8$ would mean physical nodes retain more of their states after the lumping, and for our example would result in those physical nodes having 8, and 16 final state nodes respectively.

Consider a simple illustrative lumping example in Figure \ref{fig:figure_1} which demonstrates the lumping process for a single physical node (circle of red dashed lines, $\nu_1$) and its constituent state nodes (grey circles within the red dashed circle) for different values of $k$. For the case $k = 1$ (see $\mathcal{M}'$ in middle of Figure \ref{fig:figure_1}) only the nearest neighbours of each state node are considered and as such $s_1$ and $s_2$ will be lumped together first. The next lumping of state nodes is unclear given that both $s_3$ and $s_4$ have 1-step neighbors states in different physical node. However, as we increase $k$ we explore more of the local network and at $k=2$, in this example, it becomes clear that $s_3$ is more similar to $s_1$ and $s_2$. Hence for the second lumping, $s_3$ is merged with lumped meta node $s_{1,2}$ instead of $s_4$ (see $\mathcal{M}''$ in middle of Figure \ref{fig:figure_1}).

\subsection*{Dynamical community detection}

Dynamic community detection with Markovian assumptions can still be used to reveal structure in a memory network, simply by applying the same community detection algorithms to the higher-order network structure. $\mathcal{M}_k$, for $k > 1$,
acts to constrain a walkers movement over the physical nodes within its state network connectivity. Hence, if we look for regions across $\mathcal{M}_k$ that conserve flow from a dynamical perspective, projecting the resultant communities back onto the physical nodes reveals overlapping communities constrained by the transitivity of the state network.

One such example for such a dynamical approach to community detection is Markov stability (MS) \cite{delvenne2008}, which is the focus for this study. MS exploits diffusion dynamics over an underlying graph structure to reveal a multi-scale community organisation and has been show to be effective in a variety of applications in which multiple scales are expected to exist such as protein sub-structures~\cite{peach2019unsupervised} or social behaviours~\cite{peach2019data}. Given a partition $\mathcal{P}$ of nodes into $C$ non-overlapping communities with a $N \times C$ community indicator matrix $H_{\mathcal{P}}$ the time-dependent clustered autocovariance matrix in MS is given by,

\begin{equation}
    R(t,H_{\mathcal{P}}) = H^T_{\mathcal{P}} \begin{bmatrix}  \Pi  (exp(t[M-I]) - \pi\pi^T)\end{bmatrix} H_{\mathcal{P}},
\end{equation} where the elements of the matrix  $[R(t,H_{\mathcal{P}})]$ correspond to the probability of a random walker starting at node $i$ and ending up in community $c$ at Markov time $t$ minus the probability of that happening by chance.

For an optimal partition $\mathcal{P}$, in which flow is trapped more than one would expect by random over $t$, we would expect a comparatively large Markov stability
With the Markov stability as

\begin{equation}
    r(t,H_{\mathcal{P}}) = \text{trace } R(t,H_{\mathcal{P}}) .
\end{equation}  

We aim to maximise $r(t,H_{\mathcal{P}})$ over the space of possible partitions $\mathcal{P}$ at a given Markov time $t$,

\begin{equation}
\label{eqn:MS_opt_partition}
    \mathcal{P}_{\max (t)} = \underset{\mathcal{P}}{\mathrm{argmax}} \:r(t,H_{\mathcal{P}}).
\end{equation}

Whilst the optimisation of Equation \ref{eqn:MS_opt_partition}
is NP-hard, in practice, heuristics algorithms have been developed which are computationally efficient. Here we use the Louvain algorithm which has has been demonstrated to offer robust solutions at reasonable cost \cite{blondel2008}.

\subsubsection*{Identifying partitions of interest over Markov-time}

Given a set of partitions that are optimal at each Markov time we must still define which scales are representative or robust in respect to our system. 
In order to identify partitions of interest over time we look towards two robustness measures. Firstly, we look at consistency of partitions for single points in time, and secondly, we look for stable partitions across time.

To assess this consistency between $\mathcal{P}$ at Markov time $t$ we can compute an information-theoretical distance between two alternate partitions $\mathcal{P}$ and $\mathcal{P}'$ is employed:

\begin{equation}
    VI(\mathcal{P}_i(t),\mathcal{P}_j(t)) =  \frac{2 \Omega(\mathcal{P},\mathcal{P}') - \Omega(\mathcal{P}) - \Omega(\mathcal{P}')}{ \log(n) } ,
\end{equation}

\noindent where $\Omega(\mathcal{P})$ is the Shannon entropy, $\mathcal{P}_\mathcal{C}$ being the relative frequency of finding a node in community $\mathcal{C}$ in partition $\mathcal{P}$.

Then to quantify consistency at Markov time $t$ we compute the average variation of information of all solutions:

\begin{equation}
    \langle V(t)\rangle =  \frac{1}{l-1}  \sum_{i \ne j} VI(\mathcal{P}_i(t),\mathcal{P}_j(t)).
\end{equation}

For the case that optimisations return near identical partitions $\langle V(t)\rangle$ will be small, which indicates robustness of the partition at $t$. Hence over $t$ we search for partitions with low $\langle V(t)\rangle$.

Relevant partitions should also be remain consist across regions of Markov time. Such persistence is indicated both by a plateau in the number of communities over $t$ and a low value or plateau of the cross-time variation of information:

\begin{equation}
    VI(t,t') = VI(\widehat{\mathcal{P}}(t),\widehat{\mathcal{P}}(t')) .
\end{equation}

\subsection*{Multi-scale Centralities}

For identification of central nodes we use Multiscale Centrality, that enables us to identify nodes that are central at different scales within the network \cite{PhysRevResearch.2.033104}. Multiscale centrality leverages the presence of `overshooting' peaks that appear in diffusion processes on the graphs. For a more detailed description of overshooting peaks, see \cite{peach2020semi}.
Central nodes are defined as a node, $i$ that breaks the triangle inequality for a pair of nodes $j,k$,
\[ \Delta_{ij,\tau} :=  t^*_{ij, \tau} + t^*_{ik, \tau} - t^*_{jk, \tau} \leq 0, \] .

where $t_{ij, \tau}$ is the Markov time at which an overshooting peak appears at node $j$ given the diffusive process of an initial delta function at node $i$ which is allowed to diffuse up to Markov time $\tau$.

The diffusion process underlying Multiscale centrality acts as a scaling factor that allows us to identify nodes that are central at different scales of the network structure. For example, some nodes may be locally central (with high degree) or might be globally central (high closeness). Thus we produce a ranking of nodes as a function of Markov time $\tau$ of the diffusion process. For further details on this methodology, see~\cite{PhysRevResearch.2.033104}. 

For each state node we can compute the Multiscale centrality. In an identical manner to Higher-order PageRank (see Section Higher-order PageRank), we can then compute a physical node centrality by summing the multiscale centrality over the constituent state nodes.

\section*{Abbreviations}

AMR: Antimicrobial resistance; HAI: Healthcare-associated infection; IPC: Infection Prevention and Control; CPE: Carbapenemase-producing  Enterobacteriaceae; MS: Markov Stability; MSC: MultiscaleCentrality.

\section*{Data collection and ethics}

Patient pathway data was collected from the central business intelligence system and fully pseudanonymised before analysis, in accordance with ethics 15-LO-0746.

\section*{Availability of data and materials}

The datasets generated and analysed during the current study are not publicly available to protect anonymity of included hospital patients.

The repository for Multiscale centrality can be found at \url{https://github.com/barahona-research-group/MultiscaleCentrality}.

\section*{Competing interests}

The authors declare that they have no competing interests.

\section*{Funding}

AM was supported in part by a scholarship from the Medical Research Foundation National PhD Training Programme in Antimicrobial Resistance Research (MRF-145-0004-TPG-AVISO), as well as by the National Institute for Health Research Academy. FD is supported by the Medical research council Clinical academic research partnership scheme. Professor AH is a National Institute for Health Research Senior Investigator, AH is also partly funded by the National Institute for Health Research Health Protection Research Unit in Healthcare Associated Infections and Antimicrobial Infections in partnership with Public Health England, in collaboration with, Imperial Healthcare Partners, University of Cambridge and University of Warwick. AM, RP, and MB also acknowledge funding from EPSRC grant EP/N014529/1 to MB, supporting the EPSRC Centre for Mathematics of Precision Healthcare.

The views expressed in this publication are those of the author(s) and not necessarily those of the NHS, the National Institute for Health Research, the Department of Health and Social Care or Public Health England.

\section*{Authors' contributions}

AM: performed empirical analysis of data, contributed to methodological developments, and to the writing of the paper;
RP: performed empirical analysis (MSC), contributed to methodological development of state lumping, and to the writing and revision of the paper;
AW: contributed to interpretation of results and revision of paper;
FD: contributed to interpretation of data and interpretation of results;
SM: provided study data and interpretation of data;
AH: co-supervised the project and contributed to interpretation of results;
MB: provided main project supervision and revision of paper.
All authors have read and approved the final paper.

\section*{Acknowledgements}

We wish to thank Eleonora Dyakova for help with accessing data. We would also like to thank Frankie Bolt, and Juliet Allibone for supporting the final publication process.

\section*{References}

{\footnotesize 
\bibliography{references}
\bibliographystyle{ieeetr}
}

\sectionfont{\clearpage}

\section*{Additional Files}

\subsection*{Additional file 1 --- State networks of $\mathcal{M}_1$ and $\mathcal{M}_2$.}

\begin{figure}[!htb]
    \centering
    \includegraphics[width=0.45\textwidth]{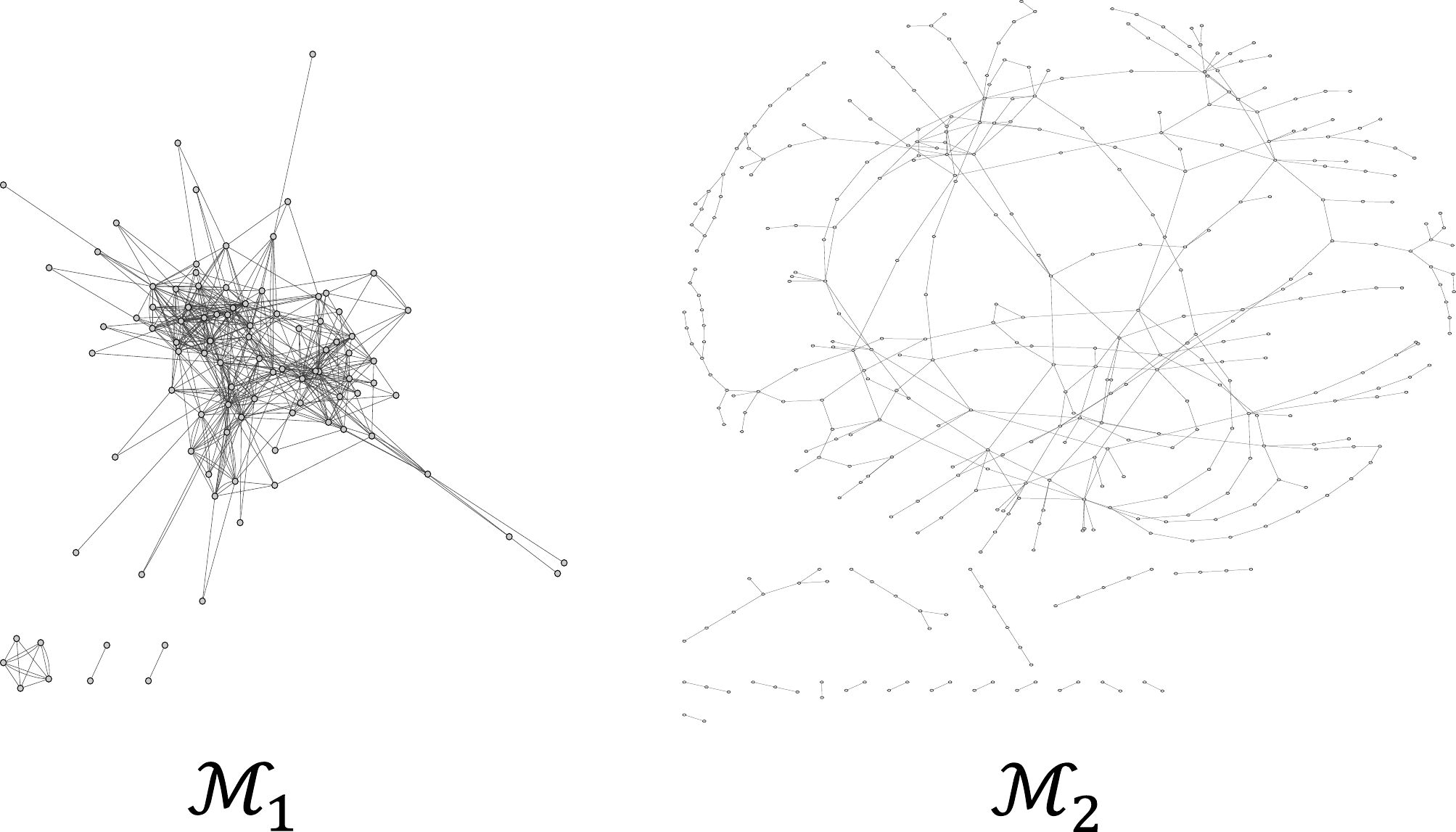}
    \caption{State networks of $\mathcal{M}_1$ and $\mathcal{M}_2$.}
    \label{fig:state_networks_m1_m2}
\end{figure}

\subsection*{Additional file 2 --- PageRank difference between $\mathcal{M}_1$ and $\mathcal{M}_2$ over specialities and buildings.}

\begin{figure}[!htb]
    \centering
    \includegraphics[width=0.45\textwidth]{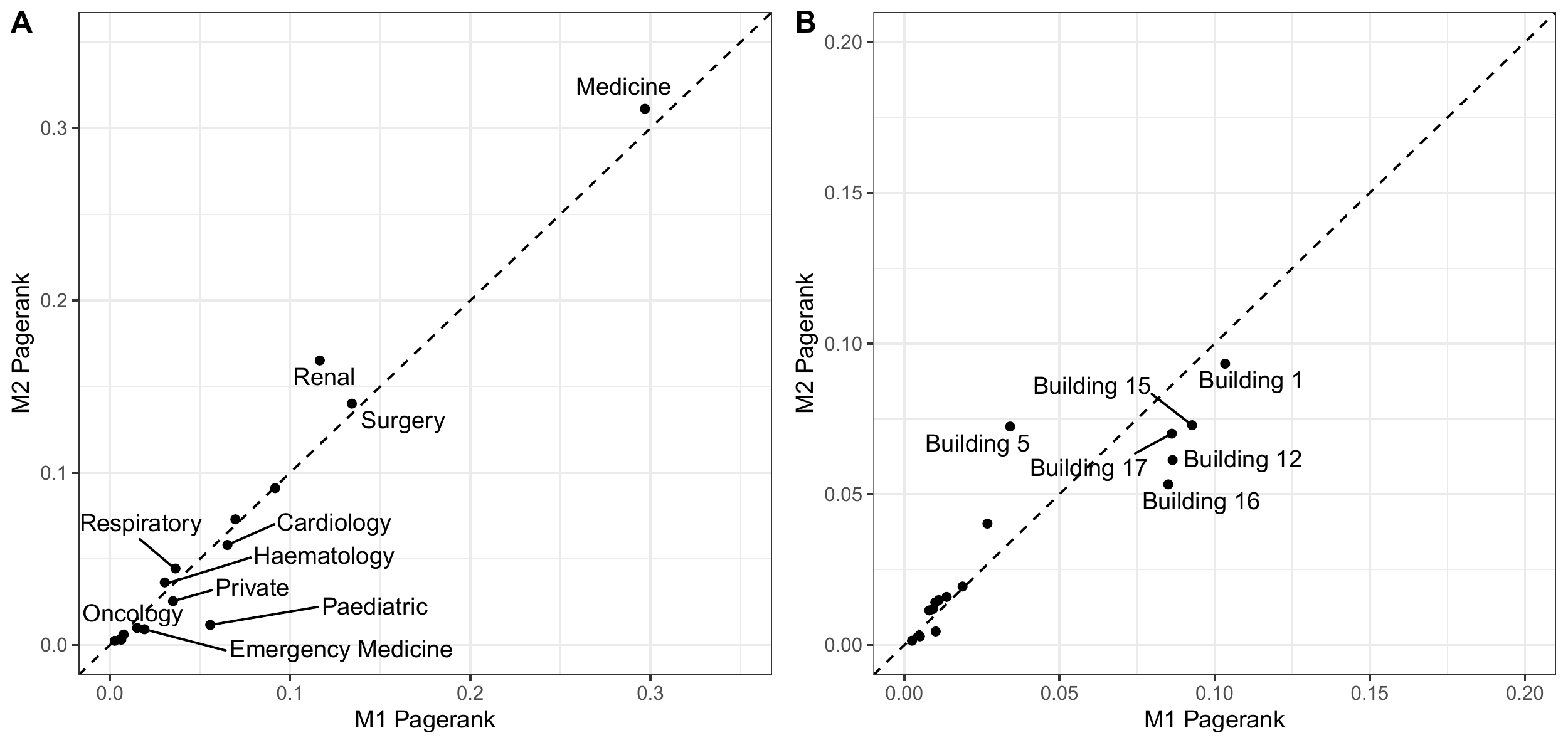}
    \caption{PageRank difference between $\mathcal{M}_1$ and $\mathcal{M}_2$ over specialities and buildings.}
    \label{fig:higher_order_validation_spec_build}
\end{figure}

Additional to analysing ward PageRanks between $\mathcal{M}_1$ and $\mathcal{M}_2$, we summed up the PageRanks of wards belonging to specialities and buildings to arrive at their visitation probabilities. Figure \ref{fig:higher_order_validation_spec_build} A \& B show the comparative results, and whilst their is less dispersion when compared to ward PageRanks, this makes sense given that specialities are more coarse groupings, and likely hide the ward variations seen previously.

\subsection*{Additional file 3 --- Optimisation of clustering rate.}

In order to select a clustering rate $r$ for lumping we investigated it's affect on (1) the number of states in the model, and (2) how well structures of patient movement can be detected in communities from the MS framework. We refer to this as model \emph{fitness}, which is aggregate amount structures (hospital sites, specialities, and buildings) found significantly \emph{over-represented} in MS communities for $t > 0.316$ (threshold in $t$ corresponding to the point of 20 partitions regardless of the clustering rate).

\begin{figure}[!htb]
    \centering
    \includegraphics[width=0.4\textwidth]{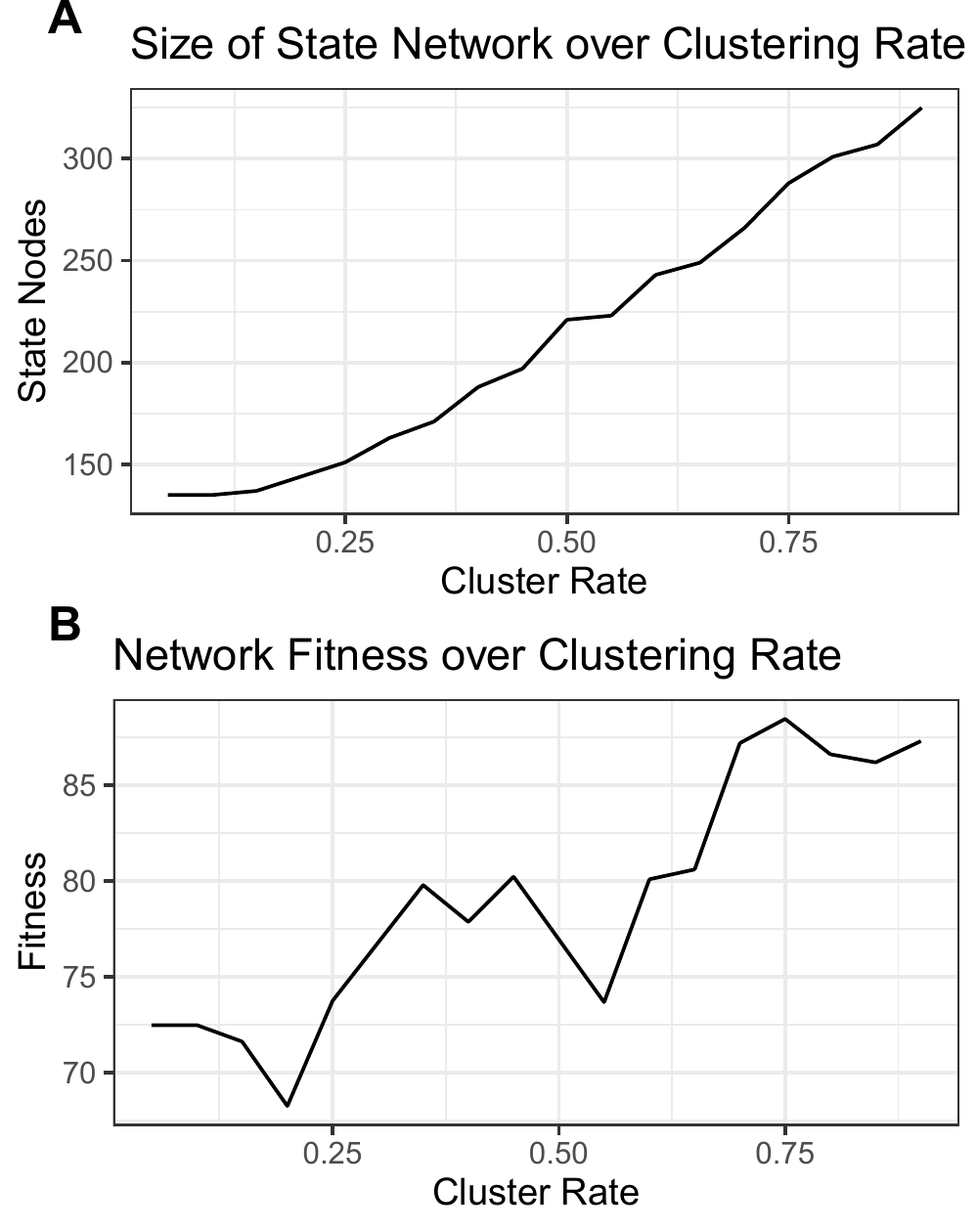}
    \caption{Optimisation of clustering rate $r$ for lumping state nodes. (A) The increasing number of state nodes (more granular lumping) with increasing lumping rate $r$, (B) the resultant model fitness as a function of the clustering rate $r$.}
    \label{fig:clustering_param_val}
\end{figure}

Figure \ref{fig:clustering_param_val}A shows the linear relationship between the number of states and $r$, whereby increases in $r$ lead to a greater number of state nodes (i.e. less lumping). Whereas Figure \ref{fig:clustering_param_val}B, the fitness curve, shows that for the same parameter range in $r$ that the model fitness does not change linearly. In fact, we observe a local peak in fitness around $r = 0.35$ whereby the total number of state nodes has reduced substantially to 171 state nodes. We hypothesise this point retains important structure of patient movement in its communities whilst  removing redundant state nodes.

\subsection*{Additional file 4 --- Markov stability run statistics.}

\begin{figure}[!htb]
    \centering
    \includegraphics[width=0.45\textwidth]{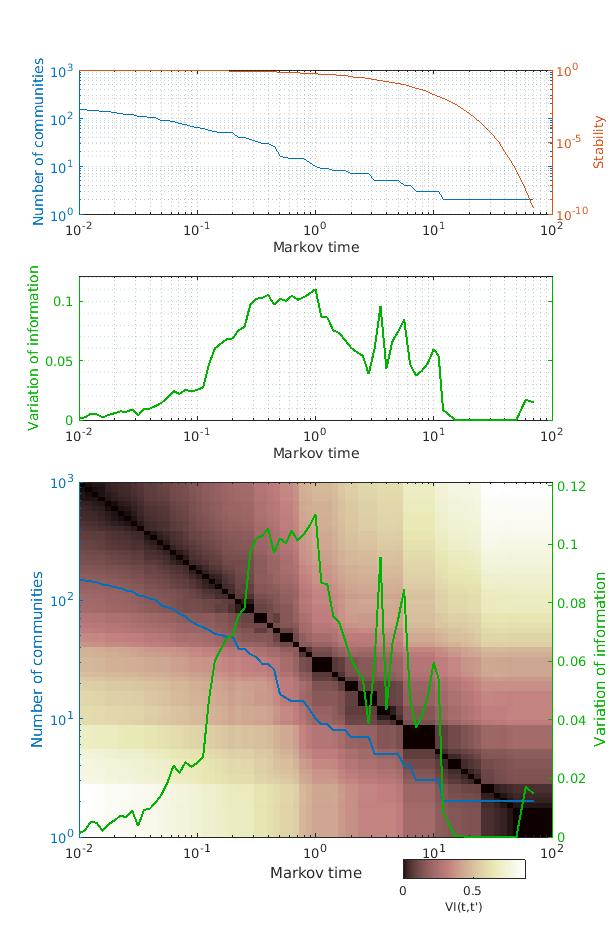}
    \caption{Markov Stability Analysis. Top: the number of communities and the Markov Stability as a function of Markov time. Middle: The Variation of Information computed over the set of Louvain optimisations at each Markov time, whereby a low VI corresponds to a robust partition. Bottom: The combined Variation of Information and number of communities. The heatmap represents the Variation of Information computed between the optimal partition at each Markov time, where the diagonal is zeros, and we look for blocks of low VI that indicate robust partitions.}
    \label{fig:MS_run_res}
\end{figure}

Figure \ref{fig:MS_run_res} shows the resultant statistics from running MS over the lumped state network $\hat{\mathcal{M}_2}$.

\subsection*{Additional file 5 --- Markov stability community partitions.}

\begin{figure}[!htb]
    \centering
    \includegraphics[width=0.4\textwidth]{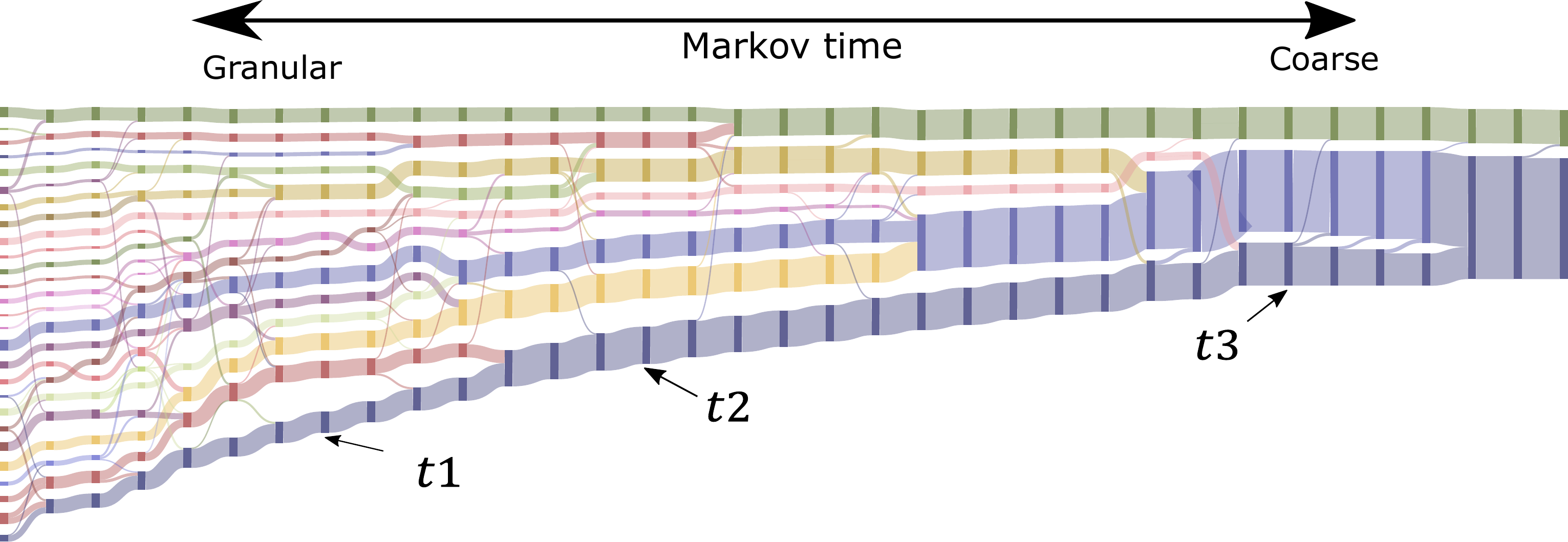}
    \vspace*{5mm}
    \caption{Sankey diagram showing full MS community Partitions over $t$ with granular partitions captured towards the left, and coarse partitions captured towards the right.}
    \label{fig:full_Markov_communties_overtime}
\end{figure}

We find that MS produces a hierarchy of community partitions across Markov time $t$. When $t$ is smaller the resultant partitions are granular, and consequently more numerous, then as $t$ increases partitions become coarse by merging granular communities together. Figure (Figure \ref{fig:full_Markov_communties_overtime}shows the full community mapping over $t$ for $\hat{\mathcal{M}_2}$, with the three time points $t1$, $t2$, and $t3$ selected that remain stable across localities in $t$, for a simplified visualisation in the main section

\subsection*{Additional file 6 --- Variation of Information between hospital structures in community partitions.}

\begin{figure}[!htb]
    \centering
    \includegraphics[width=0.45\textwidth]{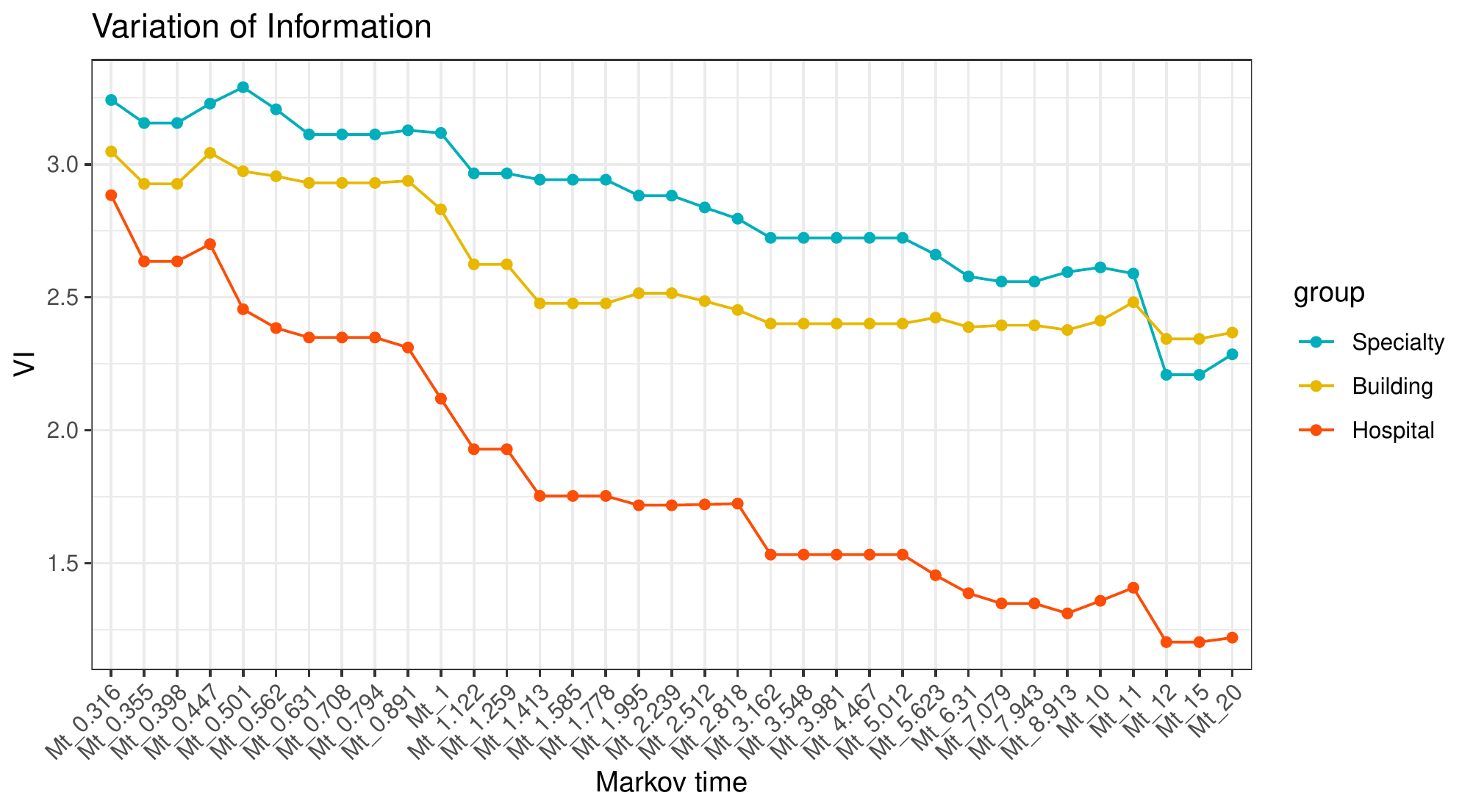}
    \caption{The Variation of Information computed between each hospital structure partition and the community partitions found at each Markov time.
    }
    \label{fig:VI_hops_structures_overtime}
\end{figure}

Similar to MS we can compute the Variation of Information (VI) to assess distance between clustering, except here we can turn to how well over $t$ the resultant partitions confer to our known structures in the hospital (sites, buildings and specialities)(Figure \ref{fig:VI_hops_structures_overtime}). As $t$ increases all structures become more aligned with MS communities, however, hospital sites seems to confer far better across $t$, even with an initial high VI the rate. Furthermore, Hospital exhibits a faster decrease rate when compared to Speciality or Buildings, and suggests that coarser communities confer most to hospital sites. However, the comparatively smaller VI for Hospitals across more granular MS communities also suggests presence of within hospital structures of patient movement, not bound solely by buildings or specialities.

\subsection*{Additional file 7 --- 2-way community partition to hospital site.}
\begin{figure}[!htb]
    \centering
    \includegraphics[width=0.25\textwidth]{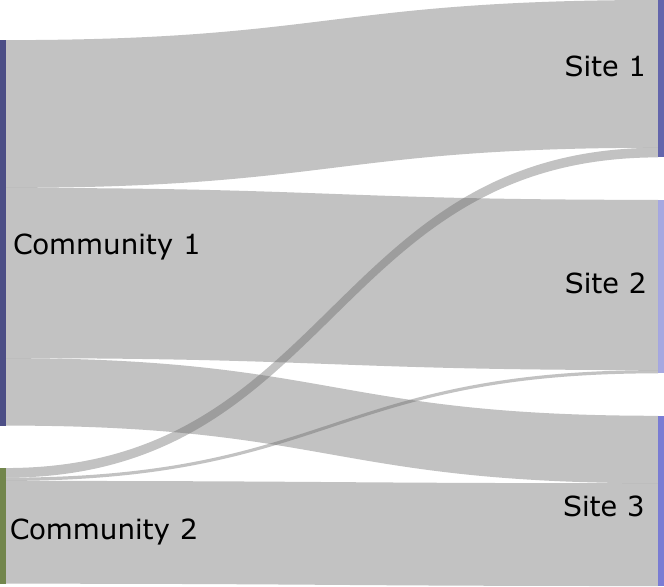}
    \vspace*{5mm}
    \caption{The Markov Stability community partition at Markov time $t = 20$ and their assignments to hospital sites.}
    \label{fig:Mt20_coms_to_sites}
\end{figure}

\subsection*{Additional file 8 --- Hospital wards overlapping communities across Markov stability partitions.}

\begin{figure}[!htb]
    \centering
    \includegraphics[width=0.45\textwidth]{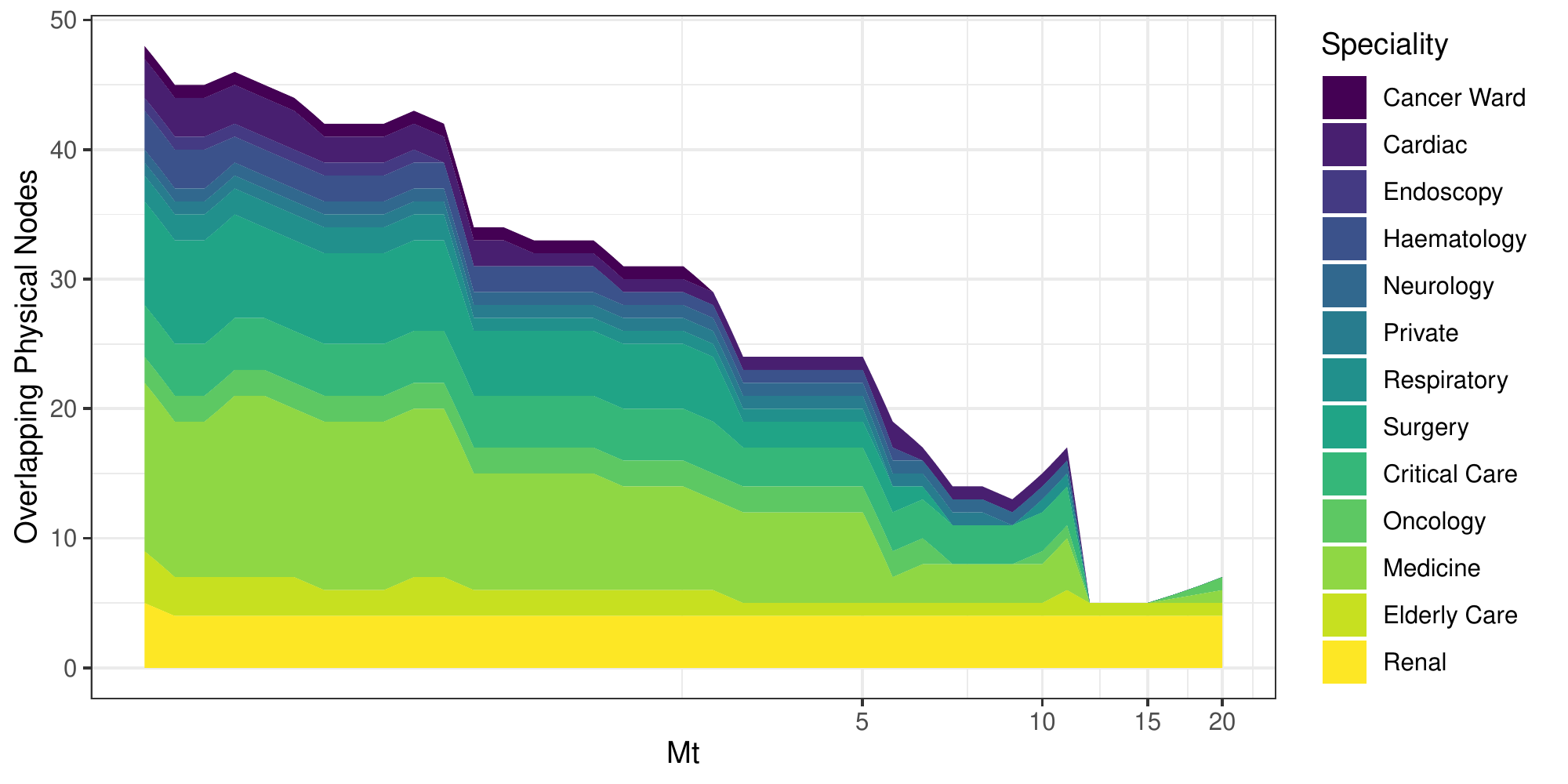}
    \caption{The frequency of physical wards that are members of more than one MS community as a function of Markov time $t$. For example, the Renal speciality has four wards that overlap between different communities for the majority of Markov time.
    }
    \label{fig:Overlapping wards}
\end{figure}

\subsection*{Additional file 9 --- Multiscale Centrality model comparison.}

\begin{figure*}[!htb]
    \centering
    \includegraphics[width=1\textwidth]{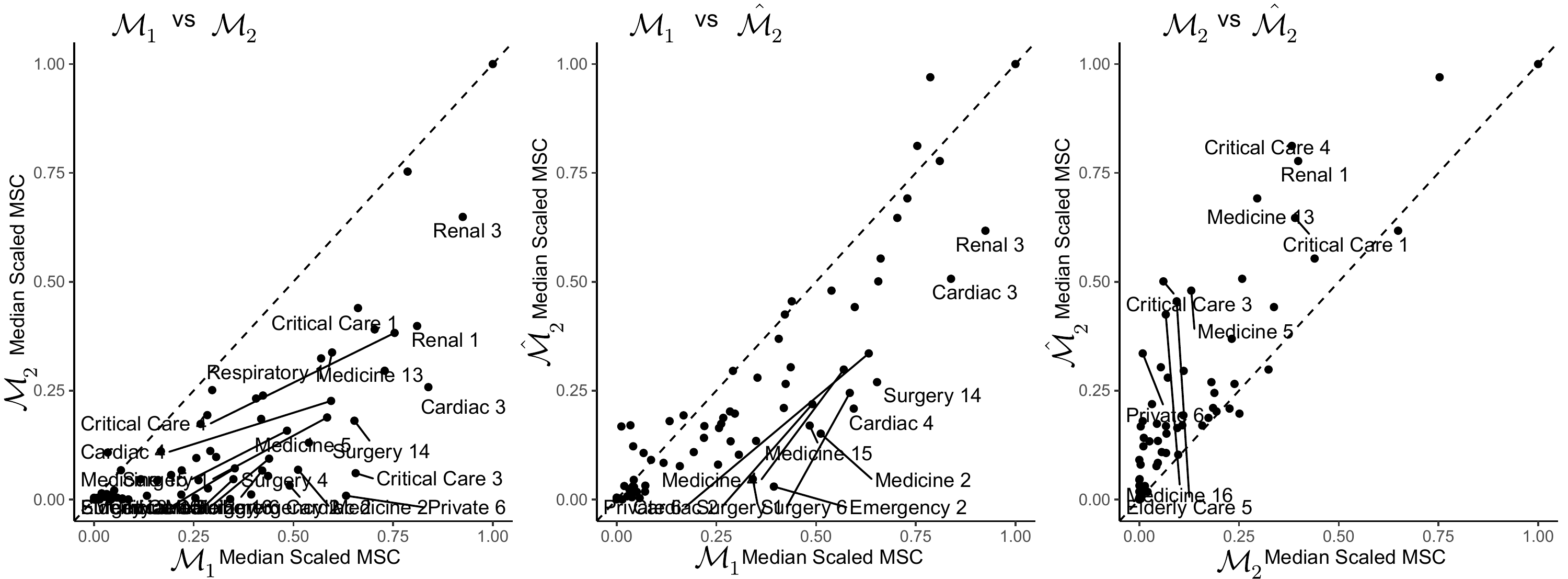}
    \caption{A comparison of the median Multiscale centrality for the first-order $\mathcal{M}_1$, second-order $\mathcal{M}_2$ and lumped $\hat{\mathcal{M}_2}$ memory networks}
    \label{fig:median_MSCs}
\end{figure*}

For further examination of the importance of higher-order modelling, we compared the MSC ranking of wards in the lumped network $\hat{\mathcal{M}_2}$ to the original state node networks of $\mathcal{M}_1$ and $\mathcal{M}_2$. We found that whilst correlated, there were a number of distinct differences between the models (Figure~\ref{fig:median_MSCs}).
 
For instance, we found several wards, including a critical care ward that were central at all time-scales in $\mathcal{M}_2$ and $\hat{\mathcal{M}_2}$ only appeared as important at short time-scales in $\mathcal{M}_1$. We found that the MSC node ranking for $\hat{\mathcal{M}_2}$ was marginally more correlated with $\mathcal{M}_1$(Ranked Cor: 0.86 (pval $<$0.01)) than $\mathcal{M}_2$Ranked Cor: 0.84 (pval $<$0.01)), which makes sense given that the state space of the lumped state network $\hat{\mathcal{M}_2}$ is closer in size to $\mathcal{M}_1$ than $\mathcal{M}_2$. 


\end{document}